\documentclass[fleqn,usenatbib]{mnras}

\usepackage{newtxtext,newtxmath}

\usepackage[T1]{fontenc}

\DeclareRobustCommand{\VAN}[3]{#2}
\let\VANthebibliography\thebibliography
\def\thebibliography{\DeclareRobustCommand{\VAN}[3]{##3}\VANthebibliography}

\usepackage{graphicx}	
\usepackage{amsmath}	
\newcommand{\corr}[1]{#1} 

\title[GIZMO-GR]{A general relativistic extension to mesh-free methods for hydrodynamics}

\author[A. Lupi]{
Alessandro Lupi $^{1,2,3}$\thanks{E-mail: alessandro.lupi@unimib.it}
\\
$^{1}$Dipartimento di Fisica ``G. Occhialini'', Universit\`a degli Studi di Milano-Bicocca, Piazza della Scienza 3, I-20126 Milano, Italy\\
$^{2}$INFN, Sezione di Milano-Bicocca, Piazza della Scienza 3, I-20126 Milano, Italy\\
$^{5}$INAF – Osservatorio di Astrofisica e Scienza dello Spazio di Bologna, Via Gobetti 93/3, I-40129 Bologna, Italy\\
}

\date{Accepted XXX. Received YYY; in original form ZZZ}

\pubyear{2022}

\begin{document}
\label{firstpage}
\pagerange{\pageref{firstpage}--\pageref{lastpage}}
\maketitle

\begin{abstract}
The detection of gravitational waves has opened a new era for astronomy, allowing for the combined use of gravitational wave and electromagnetic emissions to directly probe the physics of compact objects, still poorly understood. So far, the theoretical modelling of these sources has mainly relied on standard numerical techniques as grid-based methods or smoothed particle hydrodynamics, with only a few recent attempts at using new techniques as moving-mesh schemes. Here, we introduce a general relativistic extension to the mesh-less hydrodynamic schemes in the code \textsc{gizmo}, which benefits from the use of Riemann solvers and at the same time perfectly conserves angular momentum thanks to a generalised leap-frog integration scheme. We benchmark our implementation against many standard tests for relativistic hydrodynamics, either in one or three dimensions, and also test the ability to preserve the equilibrium solution of a Tolman-Oppenheimer-Volkoff compact star. In all the presented tests, the code performs extremely well, at a level at least comparable to other numerical techniques.
\end{abstract}




\section{Introduction}

The detection of gravitational waves (GW) by the LIGO/VIRGO collaboration \citep{abbott2016} has opened a new window on the observation of the Universe, and led to the advent of multi-messenger astronomy. By combining GW and electromagnetic emission of compact objects, we are starting now to address some fundamental and still open questions, as the behaviour of matter in the strong gravity regime, or the neutron star equation of state. From a theoretical perspective, accurately modelling the dynamics of baryonic matter around compact objects is crucial to our understanding, but at the same time it is extremely complex, because of the need of solving the General Relativity equations spanning several orders of magnitude in spatial and time scales, from the neutron stars interior (or the event horizon for black holes) to the outer edge of the accretion disc. The standard approach to tackle these problem has relied on traditional techniques as adaptive mesh refinement mesh-based \citep{gammie2003,duez2005,giacomazzo2007,mosta2014,etienne2015,cipolletta2020}, or smoothed particle hydrodynamics \citep{oechslin2002,rosswog2010,tejeda2017,liptai2019}, both with their own advantages and relative limitations. The last decades have seen an unprecedented improvement in the numerical techniques employed to study astrophysical processes, with the advent of moving-mesh and mesh-free techniques \citep{springel10,duffell2011,Gaburov2011,hopkins15}, aimed at capturing the advantages of traditional methods and overcoming their main limitations. However, while an extended literature exists on General relativistic (magneto-)hydrodynamics (GRHD) using standard techniques, the numerical solution of the GRHD equations on unstructured moving meshes has been performed only recently, in the codes \textsc{manga} \citep{chang2020} and \textsc{arepo} \citep{lioutas2022}. In this work, we introduce a GRHD extension to the code \textsc{gizmo}, which implements a mesh-free hydrodynamics scheme originally proposed by \citet{lanson08}. To date, \textsc{gizmo} has been extensively used to tackle many astrophysical problems, all based on Newtonian gravity, from galaxy formation \citep{hopkins18a,lupi19a} to the formation of star-forming clouds \citep{grudic2021,lupi2021} and pre-stellar filaments and cores \citep{bovino2019,bovino2021}, to the dynamics of gas discs around single and binary black holes \citep{sala2021,franchini2022}, to the tidal disruption of stars \citep{mainetti2017}. Here, we present our new implementation, which works with both Meshless-Finite-Volume (MFV) and Meshless-Finite-Mass (MFM) methods available in the code \citep{hopkins15}, that will enable us to employ the same numerical scheme on all scales relevant for compact object evolution. The MFV method is similar in spirit to \textsc{manga} and \textsc{arepo}, despite the completely different definition of the volume partition, which is based on a smooth field in \textsc{gizmo} instead of a Voronoi tessellation. MFM, instead, by ensuring zero mass-flux among cells, makes the numerical scheme fully Lagrangian, hence closer to the SPH approach (although with differences). The paper is organised as follows: in Section~\ref{sec:method} we describe our algorithm, in Section~\ref{sec:test} we validate our implementation, and in Section~\ref{sec:conclusions} we draw our conclusions.

\section{Numerical implementation}
\label{sec:method}
After a brief recap of the basic principles of the mesh-free methods implemented in \textsc{gizmo}, we describe the new implementations to solve the general relativistic hydrodynamic equations.

\subsection{Conservative meshless hydrodynamics}
The basic idea behind the MFV/MFM methods in \textsc{gizmo} is to partition the simulated domain via a set of discrete points/elements. Each infinitesimal volume $d\mathbf{x}$ is indeed split among these elements (`cells') according to a weight function $\psi_i$, so that 
\begin{equation}
    d\mathbf{x} = \sum_i \psi_i(\mathbf{x},h(\mathbf{x})),
\end{equation}
where $\mathbf{x}$ is the position of the infinitesimal volume, $h(\mathbf{x})$ is the kernel size, defined to encompass a desired number of neighbours, and
\begin{equation}
    \psi_i(\mathbf{x},h(\mathbf{x})) = \frac{W(\mathbf{x_i-x},h(\mathbf{x}))}{\sum_j W(\mathbf{x_j-x},h(\mathbf{x}))},
\end{equation}
with $W(\mathbf{x_i-x},h(\mathbf{x}))$ the kernel function (by default a cubic spline is employed).

Now, instead of discretising the hydrodynamic equations in the local strong conservation form, as done in the smoothed particle hydrodynamic (SPH) technique, we employ here the weak form of the equations, as done in finite-volume schemes. We start from the equations in a moving frame \citep{lanson08,springel10,hopkins15}
\begin{equation}
    \frac{d\mathbf{U}}{dt} + \nabla\cdot(\mathbf{F}-\mathbf{w}\otimes\mathbf{U})=\mathbf{S},
\end{equation}
where $\mathbf{U}$ is the state vector (consisting of mass, momentum, and energy density respectively), $\mathbf{F}$ is the flux vector, $\mathbf{w}$ the frame velocity, and $\mathbf{S}$ is a source term (for example, any external force acting on the fluid).

Following \citet{hopkins15}, we can multiply the equations by a test function $\phi$, integrate over the volume, and discretise the equations over the discrete elements in the simulated domain using the partition scheme defined above, obtaining 
\begin{equation}
    \frac{d(V_i\mathbf{U}_i)}{dt} + \sum_j[V_i\mathbf{G}_i^\alpha\tilde{\psi}_j^\alpha(\mathbf{x}_i) - V_j\mathbf{G}_j^\alpha\tilde{\psi}_i^\alpha(\mathbf{x}_j)] = V_i\mathbf{S}_i,
\end{equation}
where $V_i=\int\psi_i(\mathbf{x})d\mathbf{x}$ is the i-th particle volume, $\mathbf{U}_i$ and $\mathbf{G}_i^\alpha = \mathbf{F}_i^\alpha - \mathbf{w}^\alpha\mathbf{U}_i$ are the average state vector and the $\alpha$ component of the flux vector associated to the i-th particle respectively, and $\tilde{\psi}_j^\alpha(\mathbf{x}_i)$ is the second-order accurate gradient term giving $(\nabla f)_i^\alpha = \sum_j (f_j-f_i)\tilde{\psi}_j^\alpha(\mathbf{x}_i)$ \citep{hopkins15}.

Now, instead of taking the flux associated to each cell pair, which would require the inclusion of ad-hoc dissipation terms, as in SPH, we can replace the two flux vectors with the solution of a Riemann problem $\bar{\mathbf{G}}_{ij}$, obtaining
\begin{equation}
    \frac{d(V_i\mathbf{U}_i)}{dt} + \sum_j\bar{\mathbf{G}}_{ij}\cdot\mathbf{A}_{ij} = V_i\mathbf{S}_i,
\label{eq:FV}
\end{equation}
where $A_{ij}^\alpha=V_i\tilde{\psi}_j^\alpha(\mathbf{x}_i) - V_j\tilde{\psi}_i^\alpha(\mathbf{x}_j)$ is the normal vector to an effective `face' between the two interacting cells (which we also write as $\mathbf{A}_{ij}=A_{ij}\mathbf{n}$, with $A_{ij}$ the face area and $\mathbf{\hat{n}}$ the normal direction).

\subsection{General relativistic extension}
In what follows, we describe our implementation of a general relativistic (GR) extension of the scheme described above, which follows the method described in the \textsc{gr-hydro} code included in the Einstein Toolkit \citep{mosta2014}. In this work, we assume that the metric is static, and written in the 3+1 formalism \citep[][(ADM hereon)]{arnowitt2008} as
\begin{equation}
    ds^2 \equiv g_{\mu\nu}dx^\mu dx^\nu = (-\alpha^2+\beta_i\beta^i)dt^2 + 2\beta_i dx^i dt + \gamma_{ij}dx^idx^j,
\end{equation}
where $g_{\mu\nu}$ is the 4-metric, $\alpha$ is the lapse, $\beta^i$ the shift vector, and $\gamma_{ij}$ the spatial 3-metric. We employ here natural units, where $G=c={\rm M_\odot}=1$, and assume Einstein notation, in which repeated indices are summed over, unless otherwise stated. We will also refer to 3-dimensional quantities using Latin indices (or bold symbols) and 4-dimensional ones using Greek indices.

The equations of GR hydrodynamics are derived from the GR conservation laws, similarly to the Newtonian case. In detail, the first governing equation is the conservation of particle number, which can be written in terms of the mass density as
\begin{equation}
    \frac{1}{\sqrt{-g}}\partial_\mu(\sqrt{-g}\rho u^\mu) = 0,
\end{equation}
where $g=\det(g_{\mu\nu})$, $\rho$ is the rest mass density, and $u^\mu$ is the 4-velocity.
The other governing equations consist instead of the conservation of energy-momentum, which is defined as
\begin{equation}
    \nabla_\mu T^{\mu\nu} = 0,
\end{equation}
where $T^{\mu\nu} = \rho h u^\mu u^\nu + Pg^{\mu\nu}$ is the stress energy tensor, $h=1+u + P/\rho$ the enthalpy, $u$ the specific internal energy, and $P$ the pressure.

This set of equation can be written in hyperbolic form, similarly to the Newtonian case, where the state vector of conserved quantities is defined as $\mathcal{U} = (D,S_j,\tau)$, with the conserved variables obtained from the primitive variables $\rho,\mathbf{v},u$, and $P$ as 
\begin{equation}
   \left\{\begin{array}{l}
         D\equiv\sqrt{\gamma}\rho W\\
         S_j\equiv\sqrt{\gamma}\rho h W^2v_j\\
         \tau\equiv\sqrt{\gamma}(\rho h W^2 - P - \rho W),
    \end{array}\right.
        \label{eq:conservatives}
\end{equation}
where $\gamma=\det(\gamma_{ij})$, and $W = (1-v^iv_i)^{-1/2}$ is the Lorentz factor. The 3-velocity entering the equations $\mathbf{v}$ corresponds to the velocity seen by a Eulerian observer at rest in the spatial 3-hypersurface, and is related to the spatial components of the 4-velocity via $\mathbf{v} = \mathbf{u}/W +\boldsymbol{\beta}/\alpha$, $D$ is the conserved mass-energy density, $S_j$ is the covariant momentum density, and $\tau$ is the to the total energy density, following the convention and notation in \citet{duez2005}.
The flux vector is instead defined as
\begin{equation}
    \mathcal{F}=\left\{\begin{array}{l}
         D \tilde{\mathbf{v}}\\
         S_j \tilde{\mathbf{v}} + \alpha\sqrt{\gamma}P\mathbf{\hat{n}}\\
         \tau\tilde{\mathbf{v}} + \alpha\sqrt{\gamma}P\mathbf{v},
    \end{array}\right.
\end{equation}
where $\tilde{\mathbf{v}} \equiv d\mathbf{x}/dt = \alpha \mathbf{v} - \boldsymbol{\beta} = \mathbf{u}/u^0$.
Unlike in Newtonian hydrodynamics, where the source terms only appear when external forces are present, in GR hydrodynamics the source terms are always present, and are due to the derivatives of the metric. In particular, we have
\begin{equation}
    \mathcal{S}=\alpha\sqrt{\gamma}\left\{\begin{array}{l}
         0  \\
         \frac{1}{2}T^{\alpha\beta}\nabla g_{\alpha\beta}\\
         (T^{00}\beta^i\beta^j + 2T^{0i}\beta^j + T^{ij})K_{ij} - (T^{00}\beta^i+T^{0i})\partial_i\alpha,
    \end{array}\right.
\end{equation}
where $K_{ij}$ is the extrinsic curvature of the metric.

In order to close the system of equations, we employ here a gamma-law equation of state (EOS), i.e. $P=(\Gamma-1)\rho u$, although our implementation can be easily used with generic equations of state.

We can now replace the state and flux vectors just derived for GR hydrodynamics in Eq.~\ref{eq:FV} and naturally obtain a GR extension to the numerical scheme in \textsc{gizmo}, in the form
\begin{equation}
    \frac{d(V_i\mathcal{U}_i)}{dt} + \sum_j(\mathcal{\bar{F}}-\mathbf{w}_{ij}\mathcal{\bar{U}})\cdot\mathbf{A}_{ij} = V_i\mathbf{S}_i,
\end{equation}
where the Latin indices here refer to the interacting cells.

In the Newtonian case in \textsc{gizmo}, the Riemann problem is solved in the reference frame of the face (i.e. $\mathbf{w}=0$) and then the solution is de-boosted back to the lab frame, exploiting the Galilean invariance of the problem. In the GR case, such a boosting procedure should be based on Lorentz transformations rather than on Galilean ones, and this would make the implementation more cumbersome. For this reason, we follow the procedure in \citet{zhang2006} and \citet{chang2020}, and solve the Riemann problem directly in the lab frame. \corr{Note that in principle the Riemann problem could be solved in any frame, as long as the appropriate interval for the solution is then chosen. The reference frame of the face in the Newtonian case is typically chosen to prevent numerical approximations in the Riemann solver from potentially breaking the upwind nature of the scheme \citep[see][for details]{springel10,pakmor2011}, which can occur when the face is moving approximately at the speed of the contact discontinuity. Nonetheless, similarly accurate results have been found also when the Riemann problem is solved in the lab frame \citep[see, e.g.][]{Gaburov2011}, making the choice of the frame somewhat arbitrary.}

So far, we have not defined yet the frame velocity, which in principle can be arbitrary. Consistently with \citet{hopkins15}, in the MFV case we will assume that the face is moving with the second-order quadrature point velocity between the interacting cells, as 
\begin{equation}
    \mathbf{w}_{ij} = \mathbf{\tilde{v}}_i + (\mathbf{\tilde{v}}_j-\mathbf{\tilde{v}}_i)\frac{(\mathbf{x}_{ij}-\mathbf{x}_i)(\mathbf{x}_j-\mathbf{x}_i)}{|\mathbf{x}_j-\mathbf{x}_i|},
\end{equation}
where $\mathbf{x}_{ij} = \mathbf{x}_i + h_i/(h_i+h_j)(\mathbf{x}_j-\mathbf{x}_i)$. Nonetheless, in all the tests reported here, we found negligible difference with the simple first-order estimate $\mathbf{w}_{ij}=0.5(\mathbf{\tilde{v}}_i+\mathbf{\tilde{v}}_j)$.
As discussed in the original \textsc{gizmo} paper, this assumption is not guaranteed to capture the actual motion and deformation of the face. For this reason, we can assume that the face in reality is moving with the velocity of the contact wave, i.e. the wave for which the mass flux vanishes.

The only missing piece in our hydrodynamic scheme is the choice of the Riemann solver, which we are going to describe in the next section.

\subsection{The Riemann solver}
Similarly to the Newtonian case, an exact Riemann solver also exists for GR hydrodynamics \citep{rezzolla2003,giacomazzo2006}. Nonetheless, its complexity, together with the fact that it cannot be easily extended to generic EOSs, motivated us to opt for a simpler, but still accurate, approximate Riemann solver like the Harten-Lax-van Leer (HLL) one. Similarly to \citet{chang2020} and \citet{lioutas2022}, we here implement the HLL solver, but we also equipped \textsc{gizmo} with the HLLC solver by \citet{mignone2005}, which is able to correctly resolve contact discontinuities at a moderately higher computational cost. 
\subsubsection{The HLLC solver}
For each interacting pair, we compute the solution using a 1-dimensional Riemann solver oriented perpendicular to the face, with the solution flux determined as
\begin{equation}
    \bar{\mathbf{G}}=\left\{\begin{array}{cc}
        \mathcal{F}_L - w^{\hat{n}}\mathcal{U}_L &\quad w^{\hat{n}}<\lambda_{\rm min} \\
        \mathcal{F}^*_L - w^{\hat{n}}\mathcal{U}^*_L &\quad \lambda_{\rm min}\leq w^{\hat{n}}\leq \lambda_*\\
        \mathcal{F}^*_R - w^{\hat{n}}\mathcal{U}^*_R &\quad \lambda_*\leq w^{\hat{n}}\leq \lambda_{\rm max}\\
        \mathcal{F}_R - w^{\hat{n}}\mathcal{U}_R &\quad w^{\hat{n}}>\lambda_{\rm max} \\
    \end{array}\right.
\end{equation}
where $w^{\hat{n}}$ is the face speed $\lambda_{\rm min}$ and $\lambda_{\rm max}$ are the slowest and fastest speeds, respectively, $\lambda_*$ is the contact wave speed, and $\mathcal{F}^*_k$ and $\mathcal{U}^*_k$ correspond to the intermediate fluxes and states on the $k$ side of the contact discontinuity \citep[see][for details]{mignone2005}.

During our experiments, we found that in the case of very strong pressure jumps, the use of the HLLC Riemann solver for MFM produced large density oscillations near the contact discontinuity, which also reflected in a pressure `blip' similar to that found in SPH methods \corr{when no artificial conductivity is applied \citep{liptai2019}}, and a poorer accuracy of the numerical solution. 
For this reason, in all the tests reported in the main text, we employ the more diffusive HLL scheme (as done also in \citealt{chang2020} and \citealt{lioutas2022}), but we also report the performance of HLLC (which can be enabled in the code if desired) in Appendix~\ref{app:HLLC}. 
\subsubsection{The HLL solver}
In the HLL case, the two intermediate states are replaced by a single state 
\begin{equation}
    \mathcal{U}_{\rm HLL} = \frac{\lambda_{\rm max} \mathcal{U}_R - \lambda_{\rm min} \mathcal{U}_L + \mathcal{F}_L - \mathcal{F}_R}{\lambda_{\rm max} - \lambda_{\rm min}}, 
\end{equation}
with the flux written as
\begin{equation}
    \bar{\mathbf{G}}=\left\{\begin{array}{cc}
        \mathcal{F}_L - w^{\hat{n}}\mathcal{U}_L &\quad w^{\hat{n}}<\lambda_{\rm min} \\
        \mathcal{F}_{\rm HLL} - w^{\hat{n}}\mathcal{U}_{\rm HLL} & \quad \lambda_{\rm min}\leq w^{\hat{n}}\leq \lambda_{\rm max}\\
        \mathcal{F}_R - w^{\hat{n}}\mathcal{U}_R &\quad w^{\hat{n}}>\lambda_{\rm max} \\
    \end{array}\right.
\end{equation}
where 
\begin{equation}
    \mathcal{F}_{\rm HLL} = \frac{\lambda_{\rm max} \mathcal{F}_L - \lambda_{\rm min} \mathcal{F}_R + \lambda_{\rm max}\lambda_{\rm min}(\mathcal{U}_R - \mathcal{U}_L)}{\lambda_{\rm max} - \lambda_{\rm min}}.
\end{equation}
For MFM, this translates in assuming that the frame is moving with the corresponding zero-mass flux velocity, i.e.:
\begin{equation}
      w^{\hat{n}} \equiv \lambda_{*,\rm HLL} = \frac{\lambda_{\rm max} D_L\tilde{v}^n_L -\lambda_{\rm min} D_R\tilde{v}^n_R + \lambda_{\rm min}\lambda_{\rm max}(D_R-D_L)}{\lambda_{\rm max} D_R -\lambda_{\rm min} D_L + D_L\tilde{v}^n_L-D_R\tilde{v}^n_R},
\end{equation}
instead of $w^{\hat{n}}=\lambda_*$.
\subsubsection{Wave speed estimates}
The slowest and fastest speeds are determined as $\lambda_{\rm min}=\min\{\lambda^-_L,\lambda^-_R\}$ and $\lambda_{\rm max}=\max\{\lambda^+_L,\lambda^+_R\}$, where $\lambda^{\pm}$ is the solution of the dispersion relation, i.e.
\begin{equation}
    \lambda^\pm = \frac{(1-c_s^2)v^{\hat{n}} \pm \sqrt{c_s^2(1-v^2)[(1-v^2c_s^2)\gamma^{\hat{n}\hat{n}} -(1-c_s^2)(v^{\hat{n}})^2]}}{1-v^2c_s^2},
\end{equation}
where $v^{\hat{n}}$ is the fluid velocity perpendicular to the face, $c_s = dP/d(\rho h)$ is the relativistic sound speed (which in the case of a $\Gamma$-law EOS reduces to $c_s=\sqrt{\Gamma P/(\rho h)}$), $v^2=\gamma_{ij}v^iv^j$, and $\gamma^{\hat{n}\hat{n}}$ is the $\hat{n}\hat{n}$ component of the spatial metric, obtained via a rotation of the metric tensor \citep{chang2020}.
The contact/entropy wave velocity is instead estimated according to the Rankine-Hugoniot conditions, together with the consistency relation between energy density, pressure, and momentum density \citep[see][for details]{mignone2005}.

\subsubsection{Reconstruction at the face}
In order to achieve second-order accuracy, the states entering the Riemann problem are defined from the linearly reconstructed primitive variables at the position of the face. For this step, we employ the already available implementation in \textsc{gizmo}, suitably modified to reconstruct the rest-mass density $\rho$, the 3-velocity $\mathbf{v}$, and the pressure $P$ using the corresponding gradients.\footnote{Note that, if a generic EOS is employed, also the specific internal energy and the sound speed are reconstructed.} During reconstruction, slope-limiters are also applied to the states, in order to prevent the creation of new extrema and to make the scheme closer to be total variation diminishing \citep{hopkins15}. For the tests reported in this work, we employed the standard slope-limiter coefficients in \textsc{gizmo} (unless otherwise stated), which allow for a moderate over/under shooting of the reconstructed quantities. Although this results in some oscillations, especially in the strong shock relativistic tests (see Section~\ref{sec:test}), our goal is to show how robust and accurate the methods implemented are, despite these artefacts. As a matter of fact, the code is flexible enough that more diffusive slope limiters can be easily applied (as we will show in the 3D relativistic blast wave test).

\subsection{Conservative-to-primitive conversion and evolved quantities}
One of the most complex aspect of GR hydrodynamics is the recovery of primitive variables from conservatives, because of the presence of the Lorentz factor. Unlike in the Newtonian case, in which the accelerations and the rate of internal energy change can be easily recovered from the Riemann problem solution, in the GR case we have to rely on iterative numerical techniques as the Newton-Raphson scheme. In our implementation, we use the publicly available 2D inversion scheme by \citet{noble2006}, which also works for generic EOSs and magneto-hydro-dynamics (that we will consider in a future work) as implemented in the Einstein Toolkit, suitably modified to be integrated in \textsc{gizmo}.
Starting from a guess on the primitive variables from the previous step and the conservative quantities updated using the Riemann problems solution \citep[see][for details]{noble2006}, we can recover the current value of the primitive variables in a limited number of iterations (which do not slow down significantly the code).

Another important aspect that has to be considered is the choice of the physical quantities to evolve within the code. While in a fixed-mesh code the choice can be relatively arbitrary, in a particle-based code the resolution elements have to move in space and time, and this requires an appropriate choice of the quantities to evolve. In \textsc{gizmo}, particle positions, velocities, masses, and internal energy are commonly evolved. The simplest choice for our GR extension was then to maintain the usual meaning for particle positions and internal energy, the latter becoming now the rest-frame internal energy of the gas element. 
For what concerns velocities, instead, we found the best choice was to track $\tilde{\mathbf{v}}\equiv d\mathbf{x}/dt$ instead of the velocity of the Eulerian observer $\mathbf{v}$, which allowed us to easily drift particles during the integration. Note that this requires to apply a conversion from to/from $\mathbf{v}$ every time the Eulerian observer frame velocity is needed, but the conversion is very simple, and does not increase the computational cost significantly. Fluid velocities in GR are also subject to a strong constraint, i.e. they are limited above by the speed of light ($c=1$ in our case). Unfortunately, because of numerical approximations, in some rare cases some physical quantities might assume unphysical values. A typical case is the velocity becoming slightly larger than unity \corr{during the reconstruction at the face}. In order to avoid this pathological case, every time we compute conservative quantities, we check the validity of the Eulerian \corr{frame} velocity, and eventually rescale its value in order not to exceed a maximum user-defined Lorentz factor $W=1000$. \corr{Note, however, that these pathological situations only occur for aggressive slope limiters.} Moreover, during the conservative-to-primitive inversion, we also check when the internal energy of the gas element becomes negative (likely because of the subtraction between values with a similar magnitude), and in that case we enforce the $\Gamma$-law EOS, replacing the internal energy with $u=P/(\Gamma-1)/\rho$. \corr{One possible solution to avoid the negative internal energies is to evolve the fluid entropy rather than the total energy \citep{liptai2019}. However, such a choice would require the addition of specific dissipation terms in the presence of shocks, when entropy is not conserved, which are instead already accounted for by the Riemann solver, and would make the use of generic EOSs more difficult \citep[see, e.g.][]{werneck2022}. In the tests reported in this work, we found that these hacks were only activated for a few resolution elements in the aggressive-slope-limiter case of the 3D relativistic blast wave.\footnote{Although not reported, we repeated some of the tests in this work reconstructing the 4-velocity, which does not suffer the larger-than-unity issue, instead of the Eulerian-frame velocity, and found negligible differences in the final results.}}

Finally, particle-based codes are excellent at maintaining exact mass conservation. For this reason, and since we define the particle/cell volume using the kernel size as computed by \textsc{gizmo} enforcing the number of neighbours constraint, we decided to define the particle masses in our GR extension as the conservative mass, i.e. $m_i=D_i V_i$, instead of the rest mass. This naturally implies that our density estimator yields the conservative density $D$, and that the rest-mass density entering the pressure and the Riemann solver has to be determined by inverting Eq.~\eqref{eq:conservatives}.

\subsection{Time integration}
The time-marching algorithm in \textsc{gizmo} is a leap-frog scheme, inherited from \textsc{gadget2/3} \citep{springel05,springel08}, where velocities and internal energy are updated via two half-step kicks (one at the beginning and one at the end of each integration step), whereas positions are updated in between in full steps using the half-step values. The algorithm also allows for individual time-steps for each particle, distributing particles on a power of two hierarchy that guarantees accurate time synchronisation for interacting pairs \citep[see, also][]{springel10}. The time-step can be constrained using different conditions that depend on the process considered. In the GR case, we employ two criteria. The first criterion is the Courant-Friedrisch-Levy (CFL) condition, i.e. $\Delta t\leq C_{\rm CFL}\Delta x/v_{\rm sig}$, where $C_{\rm CFL} \leq 1$ is the CFL factor, that we set to 0.2, $\Delta x$ is the spatial resolution of the cell, that we set equal to the effective cell size $h_i=V_i^{1/3}$, and $v_{\rm sig}$ is the signal velocity, which we set similarly to \citep{liptai2019} as
\begin{equation}
    v_{\rm sig} = \max_j\left\{\frac{\bar{c}_{s} +|v^n_{ij}|}{1+\bar{c}_{s}|v^n_{ij}|}\right\},
\end{equation}
with $\max$ representing the maximum among all interacting pairs, i.e. particles enclosed within $h_i$, but also whose kernel encloses the i-th particle.
In our scheme, $\bar{c}_{s}=(c_{s,i}+c_{s,j})/2$ is the average sound speed of the pair and $v^n_{ij}=(v^n_i-v^n_j)/(1-v^n_i v^n_j)$ is the relative velocity along the line-of-sight between the two particles/cells, accounting for the relativistic composition of velocities.
The second criterion is an acceleration-based constraint, defined as $\Delta t\geq \sqrt{2\epsilon_{\rm int}h_i/a_i}$, where $\epsilon_{\rm int}=0.01$ is a tolerance parameter, and $a_i\equiv \Delta v_i/\Delta t_{\rm old}$ is the effective particle acceleration from the previous step (notice that, unlike the Newtonian case, $a_i$ is not directly computed from the momentum fluxes, but as the result of a conservative-to-primitive inversion).

Another important aspect of the time-synchronisation in \textsc{gizmo} is the value of the primitive quantities of the interacting pairs that are used in the Riemann problem. Similarly to what is done in the drift operation, in which positions are updated accounting for the half-kick velocities, all hydrodynamic quantities (for both active and inactive particles) are `predicted' forward in time to the end of the current time-step, and then used as input to the Riemann problem. This procedure is quite straightforward in Newtonian dynamics, but not so easy in GR. In order to obtain an accurate evolution, we generalised the leap-frog scheme in \textsc{gizmo} following \citet{liptai2019}, as follows:
\begin{enumerate}
    \item at the beginning of each step, we estimate the new timestep for the active particles.
    \item we estimate the conservative quantities of every active particle, and update them over a half step accounting for the source terms, using an implicit scheme \citep{leimkuhler2005,liptai2019}. Starting from a prediction 
    \begin{equation}
    \left\{
        \begin{array}{l}
           \tilde{S}_j = S_j + \mathcal{S}_{S_j}\frac{\Delta t}{2}\\
             \tilde{\tau} = \tau+\mathcal{S}_\tau\frac{\Delta t}{2} ,
        \end{array}\right.
    \end{equation}
    we recover the new primitive quantities by iterating the conservative-to-primitive solver, recomputing the source term $\mathcal{S}'$ at each iteration from the updated estimated conservative variables 
    \begin{equation}
        \left\{\begin{array}{c}
            \bar{S}_j = \tilde{S}_j + (\mathcal{S}'_{S_j}-\mathcal{S}_{S_j})\frac{\Delta t}{2}\\
            \bar{\tau} = \tilde{\tau} +(\mathcal{S}'_{\tau}-\mathcal{S}_{\tau})\frac{\Delta t}{2},\\
        \end{array}\right.
    \end{equation}
    We iterate until the maximum relative difference between two subsequent values of the momentum density $\max_j|\bar{S}_j-\tilde{S}_j|/|S|$ drops below a tolerance $\varepsilon=10^{-8}$; similarly to \citet{liptai2019}, we define the momentum magnitude as $|S| = \sqrt{\eta_{ij}S^jS^i}$, where $\eta_{ij}$ is the spatial part of the Minkowski metric.
    \item we then add the fluxes resulting from the Riemann problem solution at the previous step to the volume-integrated conserved quantities $m_i$, $\mathcal{S}V_i$, and $\tau V_i$, and determine the new primitives. 
    \item we create a new list of active particles, and update the particle positions over a full step (for all particles); in the case the underlying metric is changing (as in GR), we implicitly update the positions (by keeping the conservative quantities constant, accounting for the changes in the underlying metric). The exit conditions for the iterations is based in this case on the absolute coordinate change, i.e. $\max_j|\bar{x}_j-\tilde{x}_j|<\varepsilon$, with $\varepsilon=10^{-8}$ as above.
    \item we explicitly predict all other hydrodynamic quantities over a full step (for all particles), accounting for both the hydrodynamic fluxes and the source terms, as
    \begin{equation}
        \tilde{U}_{\rm pred}(t+\Delta t) = U_{\rm pred}(t)+(\dot{U} + \mathcal{S})\Delta t,
    \end{equation}
    \item we compute new densities, and re-estimate the gradients, for active particles only.
    \item we compute a new set of fluxes, by solving the Riemann problem between all interacting pairs.
    \item we conclude the step by applying a second kick over a half step for active particles only, using the new fluxes to update conservative variables (estimated from the half-kicked primitive ones). After this update, we estimate new primitives and compute the new source terms, which are then applied explicitly. 
\end{enumerate}
The addition of the source terms in two half-step kicks (using a Strang-split approach) guarantees second-order accuracy, as also stated in \citet{chang2020}.

\section{Tests}
\label{sec:test}
We now demonstrate the code capabilities against some standard tests for special and general relativistic hydrodynamics. We will first consider 1-dimensional tests, and then multi-dimensional ones. For the 1-dimensional tests, the exact solution we compare against is obtained using the public python package \textsc{srrp}, based on the exact Riemann solver by \citet{rezzolla2003}. We also compute the $L_2$ error on different quantities, as
\begin{equation}
    L_2 = \frac{\sqrt{(\sum_i (y_{\rm sim}-y_{\rm exact})^2)/N}}{\max_i |y_{\rm exact}|}
\end{equation}
\subsection{Special relativistic hydrodynamics}
\subsubsection{Mildly relativistic shock}
\begin{figure*}
	\includegraphics[width=\textwidth,trim=2.7cm 1cm 2.7cm 2cm,clip]{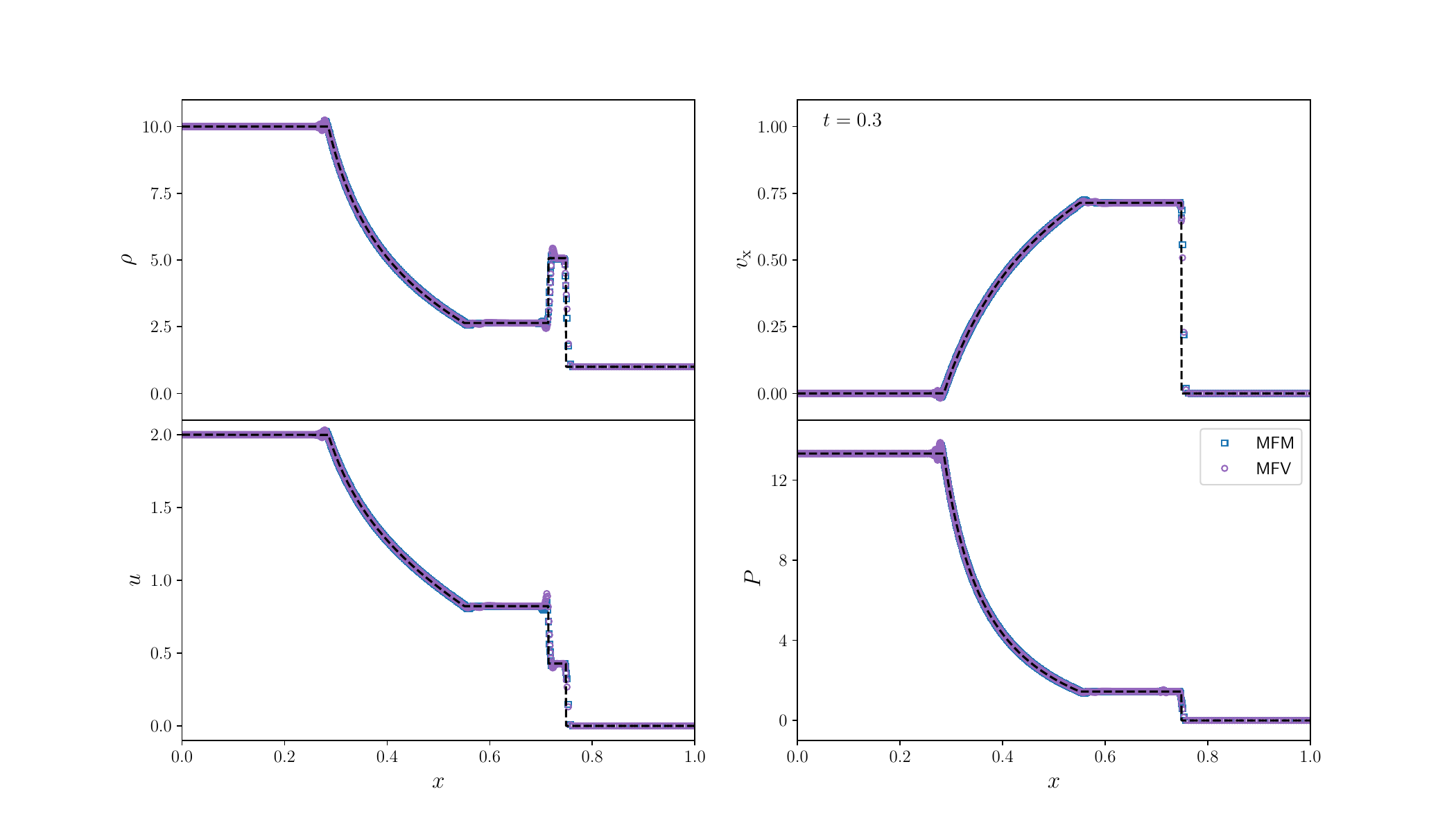}
	\caption{Mildly relativistic shock in 1D, from \citet{marti2003}, at $t=0.3$. We show the rest-mass density $\rho$ (top-left panel), $v_x$ (top-right panel), $u$ (bottom-left panel), and $P$ (bottom-right panel), for MFM (blue squares) and MFV (purple circles), compared with the exact solution, shown as a dashed black line.}
    \label{fig:marti1}
\end{figure*}
The first, and easiest test for a special relativistic (SR) code is a mildly relativistic shock, corresponding to `Problem 1' in \citet{marti2003}, with a maximum Lorentz factor $W=1.38$.
We simulate a periodic box filled with 1100 equal mass particles distributed homogeneously in the range $x\in [0,1]$ (1000 on the left half, and 100 on the right), setting 
\begin{equation}
    [\rho,P]=\left\{\begin{array}{cc}
         [10,40/3] & x<0.5 \\
         \left[1,10^{-6}\right] & x\geq 0.5\\
    \end{array}\right.
\end{equation}
In Fig.~\ref{fig:marti1}, we show the rest-mass density $\rho$ (top-left), velocity $v_x$ (top-right), specific internal energy $u$ (bottom left), and pressure $P$ (bottom right) at $t=0.3$. The MFV case is shown with purple circles, the MFM one with blue squares, and the exact solution as a dashed black line. The numerical results almost perfectly overlap with the exact solution in both cases, with only a small jump in density at the contact discontinuity (underdensity on the left, and overdensity on the right), which is also reflected in the internal energy plot. This `blip' is the natural consequence of the moving nature of the scheme, that exhibits a lower dissipation relative to fixed grids, and in which small errors in the particle motion directly reflect in the smoothed density distribution.\footnote{In order to check whether this small error was more severe in the GR case than in the Newtonian one, we also performed a shock tube test with a strong pressure jump ($P_{\rm L}/P_{\rm R}=100$)  with the default Riemann solver in \textsc{gizmo} and the exact one in \textsc{arepo}, finding the same feature in both cases.} 
A small overshooting is also visible around the end of the rarefaction fan, for $\rho$, $u$, and $P$, which however can be easily removed by applying more diffusive slope-limiters. The largest error in our solution is in $v_x$, with a value around $4\times 10^{-2}$ for both MFM and MFV.
Although not reported, we also performed a simulation with unequal mass particles using MFV, finding almost identical results (the only difference being the density/internal energy jump at the contact discontinuity, which is significantly affected by the rearranging of the particle mass/position).

\subsubsection{Relativistic blast wave}
\begin{figure*}
	\includegraphics[width=\textwidth,trim=2.7cm 1.1cm 2.7cm 2cm,clip]{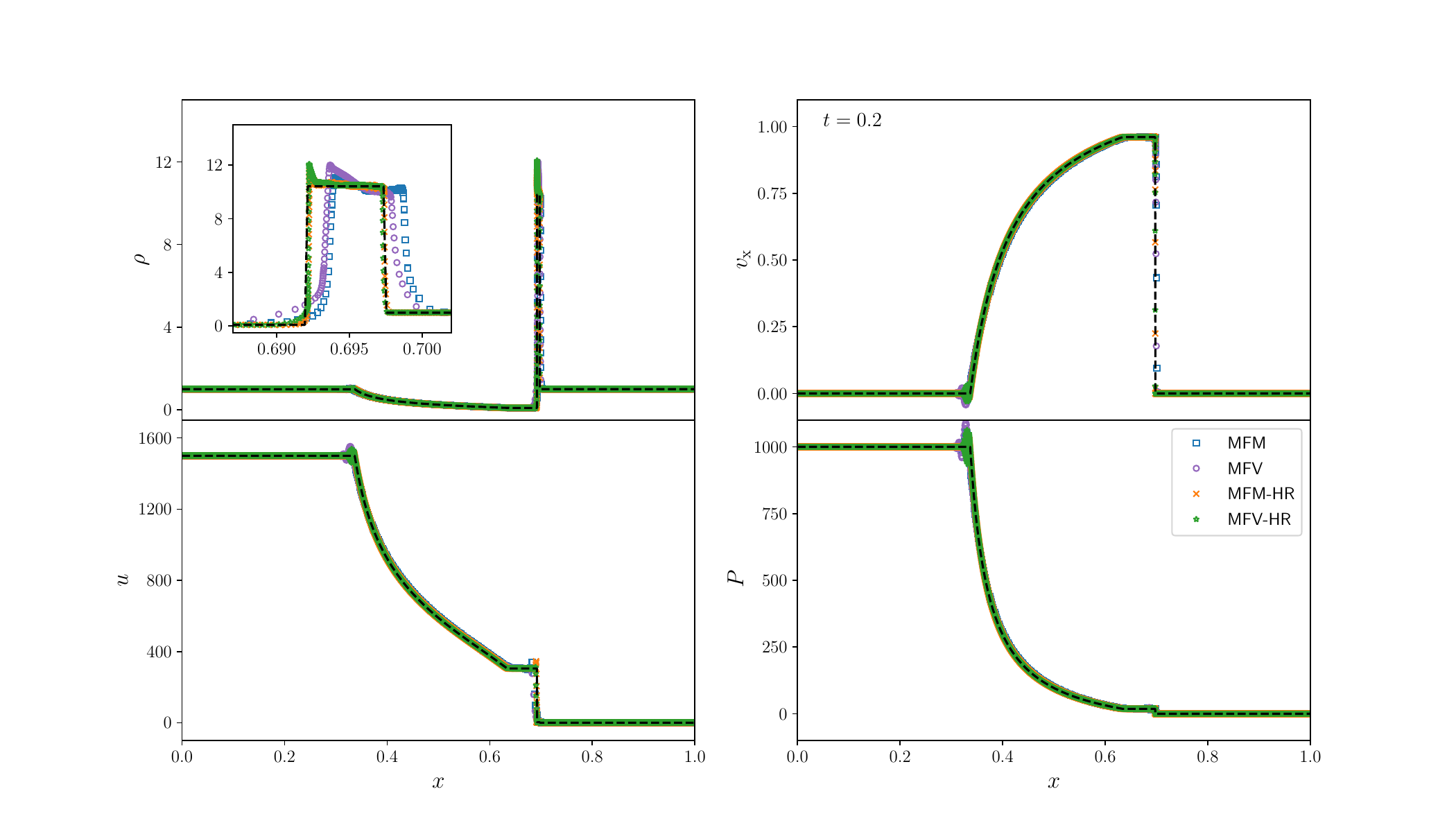}
	\caption{Same as Fig.~\ref{fig:marti1}, for the 1D relativistic blast wave problem at $t=0.2$. In addition to the fiducial `low' resolution cases, we also report two `high' resolution ones with 8000 particles, to highlight the improvement in capturing the plateau at the top of the density spike (in the inset).}
    \label{fig:marti2}
\end{figure*}

The second test we consider is a strong blast wave with a maximum Lorentz factor of 3.6, corresponding to `Problem 2' in \citet{marti2003}. This problem is much more challenging, because of the strong pressure jump. In this case, we assume a periodic box filled with 1000 equal mass particles in the interval [0,1] (500 on each side of the interface), with initial conditions
\begin{equation}
    [\rho,P]=\left\{\begin{array}{cc}
         [1,1000] & x<0.5 \\
         \left[1,10^{-2}\right] & x\geq 0.5\\
    \end{array}\right.
\end{equation}
Our results are shown in Fig.~\ref{fig:marti2}, using the same style of Fig.~\ref{fig:marti1}, at $t=0.2$. Although the agreement looks good also in this case, the density spike is too thin to show any deviation when the entire box is considered. For this reason, in the inset we report a zoom of the interval around $x=0.695$. From the comparison with the exact solution, it is clear that the simulations can reasonably capture the height of the spike, whereas the location is mildly ahead of the exact one, i.e the shock speed is moderately overestimated. This is due to the quite low resolution, and the fact that neighbouring particles interact over a significant portion of the spike. In order to show how the results improve when we increase the resolution, we also report in the figure two simulations employing 8000 particles each, as orange crosses (MFM-HR) and green stars (MFV-HR). The agreement of the density spike with the exact solution is now much better, with only a small density overshooting at the contact discontinuity, but at the same time a very well defined plateau perfectly overlapping with the location of the spike. All other quantities, instead, exhibit an almost perfect agreement even at low resolution.
The largest error is again in $v_x$ for both MFV and MFM, with $L_{2}\sim 1.4\times 10^{-1}$. At higher resolution, the largest error is found for $\rho$ in both schemes, and is associated to the moderate under/overshooting near the contact discontinuity, with a value $L_2\sim 3\times 10^{-2}$ for both schemes.

\subsubsection{Relativistic blast wave with transverse velocity}
\begin{figure*}
	\includegraphics[width=\textwidth,trim=2.7cm 1cm 2.7cm 2cm,clip]{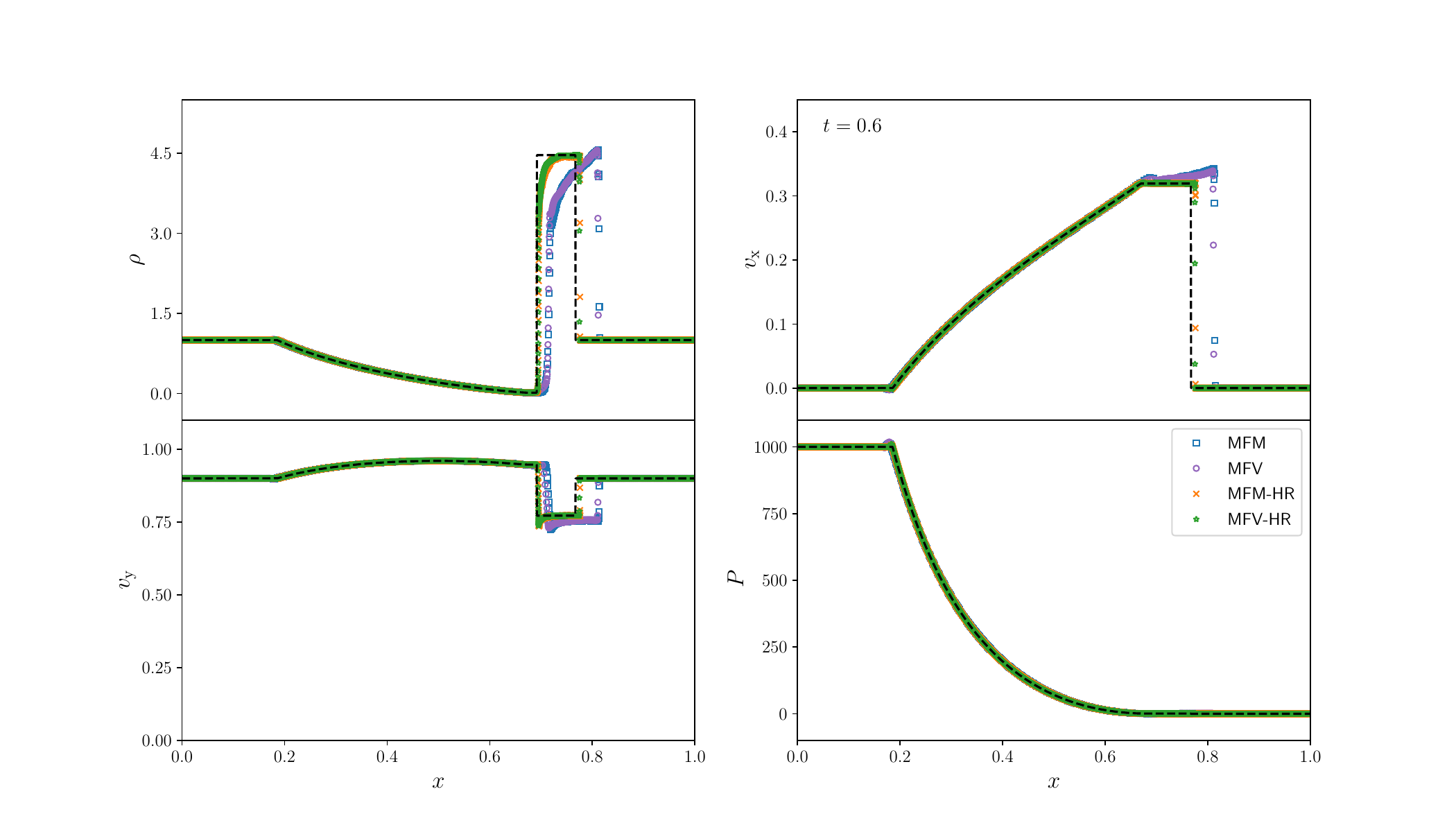}
    \caption{Same as Fig.~\ref{fig:marti2}, for the relativistic blast wave with transverse velocity, at $t=0.6$. Also in this case, we report both fiducial runs with 1000 particles, and two higher resolution cases with 8000 particles each.}
    \label{fig:marti_tr}
\end{figure*}

In this test, we add a constant transverse velocity to the relativistic blast wave discussed before. Because of the Lorentz factor that enters in the definition of the conservative variables, the transverse velocity is able to significantly modify the solution, and represents an important challenge for numerical codes, because of the complicated structure developed along the transverse direction that requires a very high resolution \citep{zhang2006}. The initial conditions are the same of the previous test, with the addition of a velocity along the $y$-direction $v_y=0.9$ across the entire domain. The results are shown in Fig.~\ref{fig:marti_tr} using the same style as above (the only difference is in the bottom-left panel, where we show now the transverse velocity). In this case, the pressure is the only quantity still showing an almost perfect agreement with the exact solution, while the other quantities advance at a faster pace, because of the overestimated shock velocity. Similarly to the previous test, we performed here two simulations at higher resolution, employing 8000 particles each. These high-resolution runs show that the code is able to almost perfectly recover the exact solution, except for a small underestimation of the density in the thin shell (especially around the contact discontinuity). The shock velocity is very close to the correct one, and, interestingly, we do not observe any large spike in $v_y$ at the contact discontinuity, even at low resolution, unlike that found, for instance, by \citet{zhang2006} and \citet{liptai2019}. The largest error is again in $v_x$, reaching $\sim 4.2\times 10^{-1}$ in the fiducial runs, and $\sim 1.8\times 10^{-1}$ in the higher-resolution ones.

\subsubsection{3-dimensional blast wave}
\begin{figure*}
	\includegraphics[width=0.9\textwidth,trim=0cm 1cm 0cm 1cm,clip]	{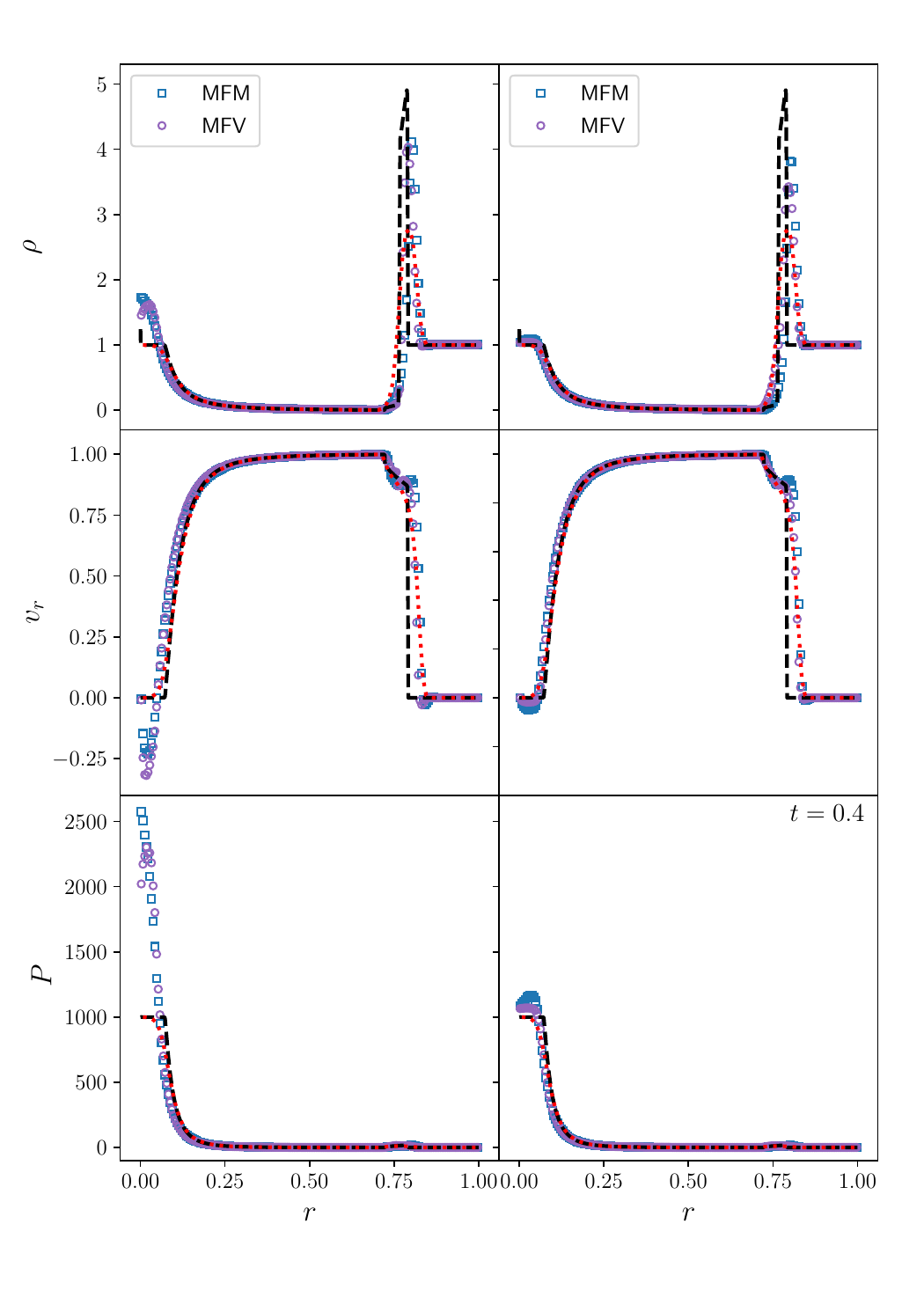}
    \caption{Radial profiles of $\rho$, $v_r$, and $P$ for the 3D relativistic blast wave test, at $t=0.4$. The MFM run is shown as blue squares, and the MFV one as purple circles. The 1D high-resolution \textsc{athena++} solution is shown as a black dashed line, while the low-resolution case is reported with a red dotted one. The left-hand panels correspond to the aggressive slope limiters case, whereas the right-hand ones to the more diffusive slope limiters case.}
    \label{fig:blast3d}
\end{figure*}
\begin{figure*}
	\includegraphics[width=\textwidth,trim=2.2cm 0.2cm 1cm 0cm,clip]{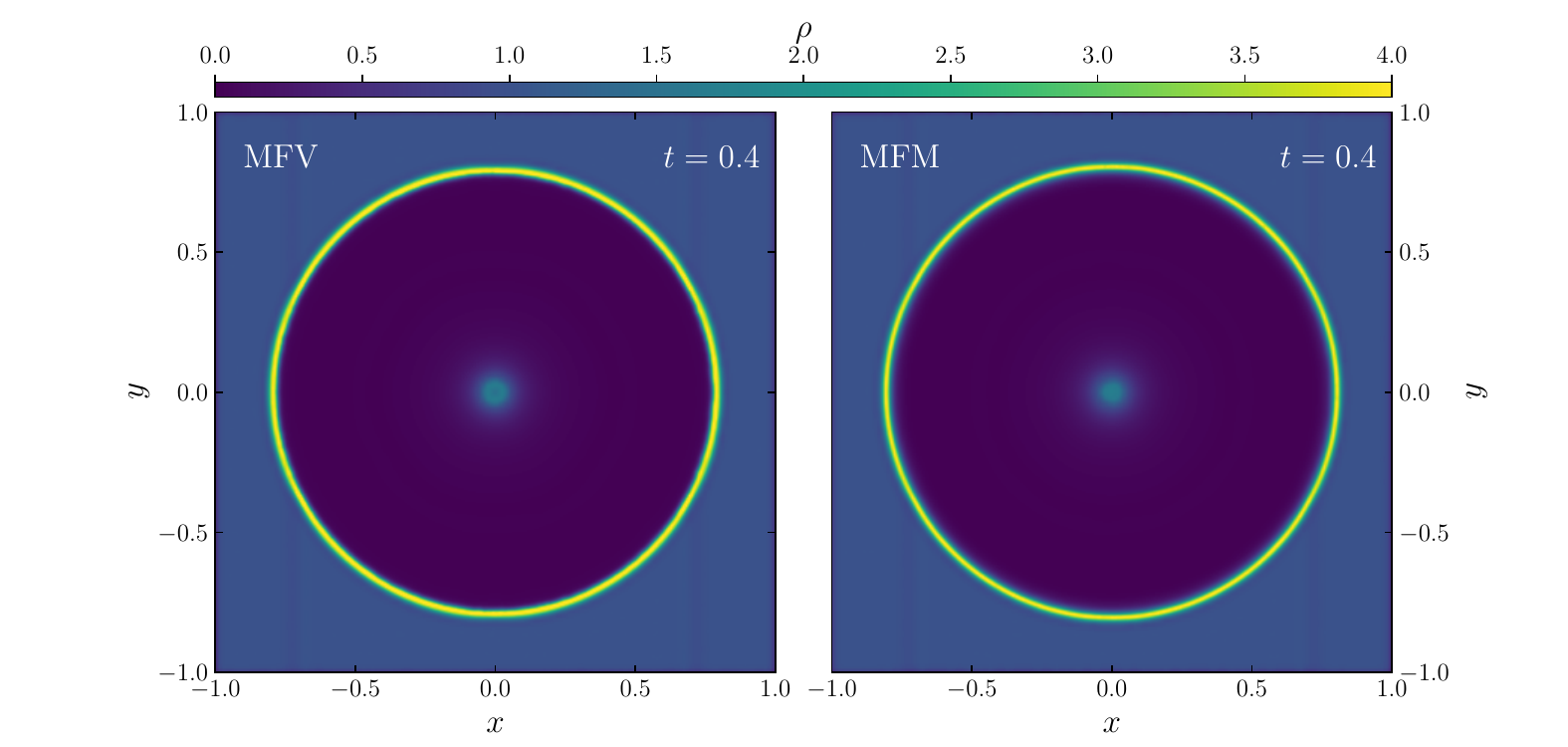}
    \caption{Slice through $z=0$ of the rest mass density in the MFV (left-hand panel) and MFM (right-hand panel) simulation at $t=0.4$.}
    \label{fig:blast3d_slice}
\end{figure*}
After having verified the code capabilities in the 1-dimensional case, we now consider a 3D version of the relativistic blast wave. The initial conditions are the same reported by \citet{zhang2006}, except for the simulated domain, which covers the entire sphere and not only an octant \citep[see, also][]{liptai2019}. In particular, we employ a periodic box of length 2, filled with a uniform density gas with $\rho=1$, sampled with $\sim 200$ particles per side distributed in a close packed lattice. The central region of the box, up to $r_0=0.4$ from the centre, is overpressurised ($P_{\rm in}=1000$), assuming the outer region is at $P_{\rm out}=1$. Unlike \citet{liptai2019}, we do not need to smooth the initial pressure discontinuity in our test to avoid negative pressure, as the Riemann solver's numerical diffusion naturally smooths out the discontinuity. \footnote{Although not reported, we also performed the same test using the smoothed profile in \citet{liptai2019}, where
\begin{equation}
    P(r) = \frac{P_{\rm in}-P_{\rm out}}{1+\exp(\frac{r-r_0}{\Delta r})}+P_{\rm out},
\end{equation}
with $\Delta r=0.01$ the typical particle separation in the initial conditions, and found negligible differences with our fiducial results.} In the left-hand panels of Fig.~\ref{fig:blast3d}, we report the radial profiles of $\rho$, $v_r$, and $P$, binned in 200 radial bins, for MFM (blue squares) and MFV (purples circles) at $t=0.4$, and compare them with the 1D solution obtained in spherical coordinates with \textsc{athena++} \citep{white2016,stone2020}, reported as a black dashed line (using 4000 cells), and a red dotted one (using 100 cells).
Our results show a qualitatively good agreement with the expected solution (the high-resolution \textsc{athena++} case), with the thin shell lying only moderately ahead of it, especially for MFM (the same is found in the equivalent low-resolution run performed with \textsc{athena++}). This is also reflected in the radial velocity profile, where the MFM/MFV runs predict a slight overshooting of the shock velocity. Our results are in line with the low-resolution grid run, but for the height of the density peak, which in our case is closer to the expected solution, even at such a low resolution, thanks to the lower diffusivity of our scheme. The only significant discrepancy in our runs is found near the centre, where the linear reconstruction and the slope limiters employed produce some oscillations of the central pressure at the end of the rarefaction fan, which propagate inward over time. Because of the relatively low resolution, around the end of the simulation ($t\geq 0.35$) the number of resolution elements in the overpressurised region becomes small ($\sim 10$ resolution elements per side, with the interaction among cells occurring over $\sim N_{\rm ngb}^{1/3}\approx 3.2$ elements on average), and this produces the pressure bump observed at $r<0.1$, also reflected in the negative radial velocity. Although increasing the resolution can surely help in alleviating this issue, a simpler solution consists in employing more diffusive slope limiters, that suppress by construction the oscillations. The results obtained in this case are showed for comparison in the right-hand panels of Fig.~\ref{fig:blast3d}. While no relevant differences are observed in the underdense region and in the shell, the central region now exhibits a much better agreement with the high-resolution results, with only a very moderate overshooting of the central pressure.

By looking in detail at the density of the shell in the two methods we implemented, we note that MFV tends to stay slightly closer to the expected solution, but spreads the shell over a slightly larger region (reaching a moderately lower peak), whereas MFM, which lies a bit ahead, is able to maintain a thinner shell. This difference can be better observed in Fig.~\ref{fig:blast3d_slice}, where we report a slice through the $z=0$ plane of the rest-mass density in our simulations (with the more aggressive slope limiters). Both methods perfectly maintain spherical symmetry and resolve the thin shell, even at our `low' resolution (less than 1/3 in linear size compared to \citealt{zhang2006} and \citealt{liptai2019}), although with the differences we just described.

\subsection{GR dynamics for test particles}
We now test the ability of the code to follow test particle dynamics in arbitrary geometries. In particular, we consider here the Schwarzchild and Kerr metrics, that represent the solution of the Einstein equations for a non-spinning ($a=0$) and spinning ($a \neq 0$) black hole. Although the two metrics are more easily written in spherical and Boyer-Lindquist coordinates respectively, in this work we implement them in Cartesian coordinates \citep[following][]{liptai2019}, as these represent the natural coordinates of our mesh-less technique. For the following tests, we will consider non-interacting particles, i.e. we switch off all the hydrodynamics part of the code, leaving only the Kick-Drift-Kick part with the source terms.\footnote{Note that, while in MFM the KDK scheme guarantees almost exact conservation of angular momentum, with MFV the potentially different velocity of the fluid relative to the particle introduces a small non-conservation error, which however does not significantly impact the solution, as we show in this section.} As mentioned above, the source terms depend on the gradients of the metric along the three spatial directions (the time derivatives of the metric can be replaced using the metric evolution equation in the ADM formalism \citealt{arnowitt2008}) contracted with the momentum-energy tensor. Although the analytic derivatives of the Schwarzchild metric terms are simple to implement, we opted for a more general approach which can be easily applied to more complex metrics (or even to a time-dependent metric), i.e. we compute all derivatives numerically via central differencing, as
\begin{equation}
    \partial_i g_{\mu\nu} \approx \frac{g_{\mu\nu}(x^j+\epsilon\delta_i^j)-g_{\mu\nu}(x^j-\epsilon\delta_i^j)}{2\epsilon},
\end{equation}
where $\epsilon=10^{-8}$, $x^j$ are the particle coordinates, and $\delta_i^j$ is the Kroneker delta.
For all the tests reported here, we force the code to maintain a timestep $\Delta t \sim 0.01M$ over the entire integration.
For all the following tests, we define the radial coordinate as in Boyler-Lindquist coordinates, i.e. 
\begin{equation}
    r=\sqrt{\frac{R^2-a^2+\sqrt{(R^2-a^2)^2+4a z^2}}{2}},
\end{equation}
which reduces to $r=R$ for $a=0$, where $R=\sqrt{x^2+y^2+z^2}$.
\subsubsection{Orbits in Schwarzchild and Kerr metric}
\begin{figure}
	\includegraphics[width=\columnwidth]{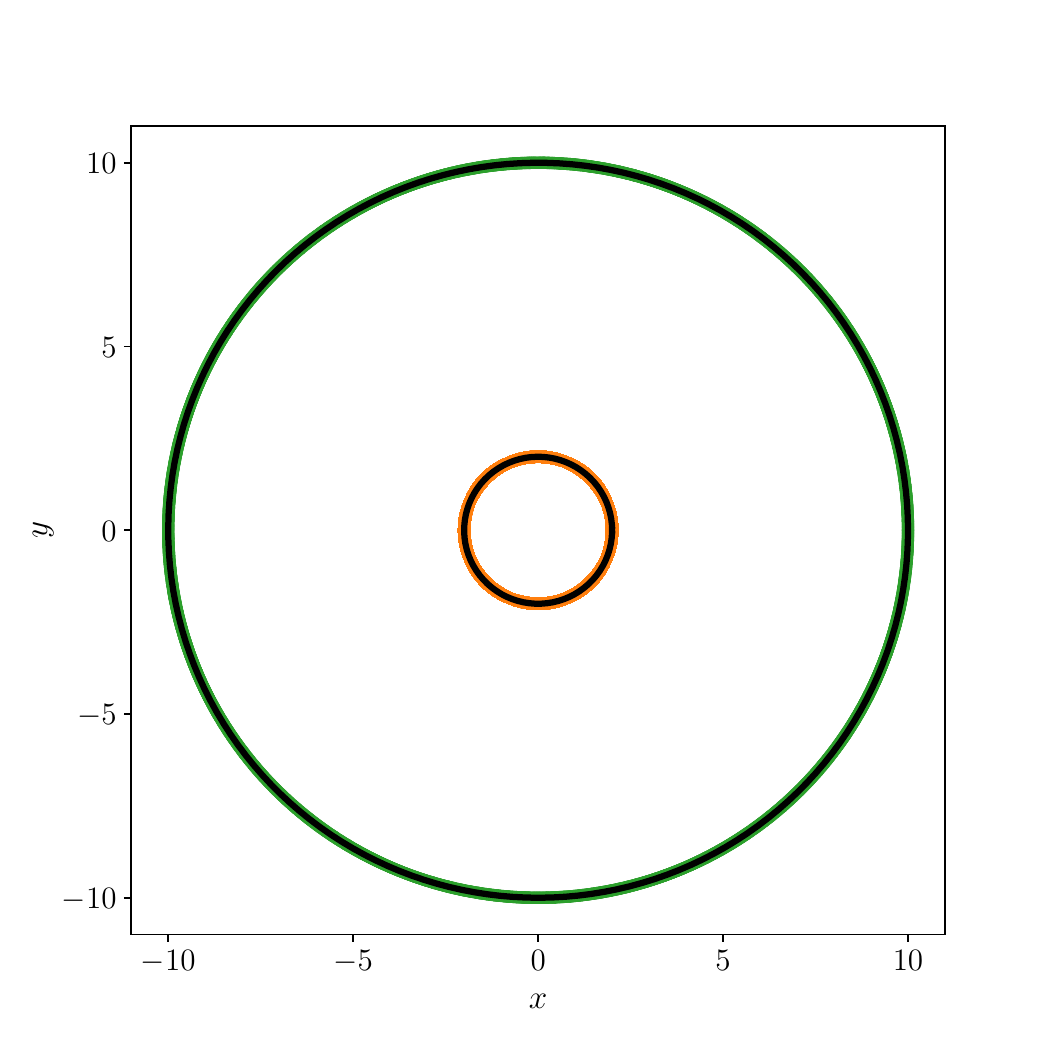}
	
    \caption{Circular orbits of test particles in our code. The black circles correspond to the analytic solutions at $r=10$ (for $a=0$) and $r=2$ (for $a=1$), whereas the green and orange curves represent the numerical solution in the two cases, respectively.}
    \label{fig:orbit_schw_kerr}
\end{figure}
The first and simplest dynamics test is the integration of a circular orbit. Both Schwarzchild and Kerr metrics admit circular orbits outside the innermost stable circular orbit (ISCO) radius, which corresponds to 6$M$ for $a=0$, $M$ for $|a|=M$ and a corotating particle, and $9M$ for $|a|=M$ and a counterrotating particle.
In the case $a=0$, we placed 32 particles at $r=10M$, equally spaced in the azimuth coordinate $\phi$, setting $v_\phi = \Omega r$, where the orbital frequency is $\Omega=M/r^3$ (identical to the Newtonian case), and evolved them for 15 orbits. The results are reported as green curves in Fig.~\ref{fig:orbit_schw_kerr}, with the black circle corresponding to the analytic solution. The error in the integration never exceeds $L_2\sim 10^{-7}$. We then repeated the same test for a maximally rotating black hole ($a=1$), placing 32 particles at $r=2$ on a corotating orbit. The orbital frequency in the case of a rotating black hole is now defined as $\Omega=M^{1/2}/(r^{3/2}+aM^{1/2})$ \citep{abramowicz1978}. The results are shown as orange curves in Fig.~\ref{fig:orbit_schw_kerr}, with the black circle corresponding again to the analytic solution. Also in this case, our code is able to perfectly maintain the circular orbit, with the error never exceeding $L_2\sim 10^{-5}$.

\subsubsection{Free-falling particle in the Schwarzchild metrics}
\begin{figure}
	\includegraphics[width=\columnwidth]{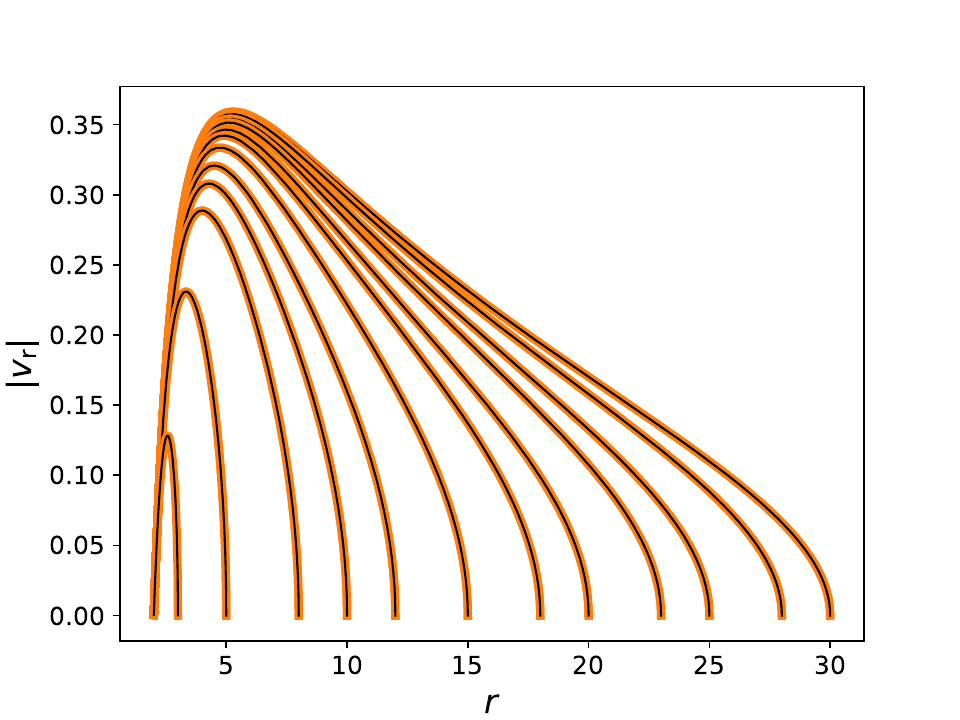}
    \caption{Free-fall dynamics in the Schwarzchild metric, starting from 12 different initial distances from the black hole in the range $[3;30]$.}
    \label{fig:freefall}
\end{figure}

In the Schwarzchild metric, a particle initially at rest free-falling onto the black hole is expected to reach a velocity
\begin{equation}
    v_r=\frac{1-2M/r}{\sqrt{1-2M/r_0}}\sqrt{2M(\frac{1}{r}-\frac{1}{r_0})},
\end{equation}
where $r$ is the current distance from the origin and $r_0$ is the initial distance.
In Fig.~\ref{fig:freefall}, we show our numerical results for 24 particles placed at 12 different distances from the black hole in the range $r\in [3;30]$, and 2 different initial values of $\phi=0,\pi$. The analytic solution is reported in black, whereas our results are shown as orange lines. Our numerical scheme is able to perfectly capture the particle dynamics, with the error reaching the highest value $L_2\sim 4\times 10^{-4}$ at $r_0=3$.
\subsubsection{Apsidal precession in the Schwarzchild and Kerr metrics}
\begin{figure}
	\includegraphics[width=\columnwidth]{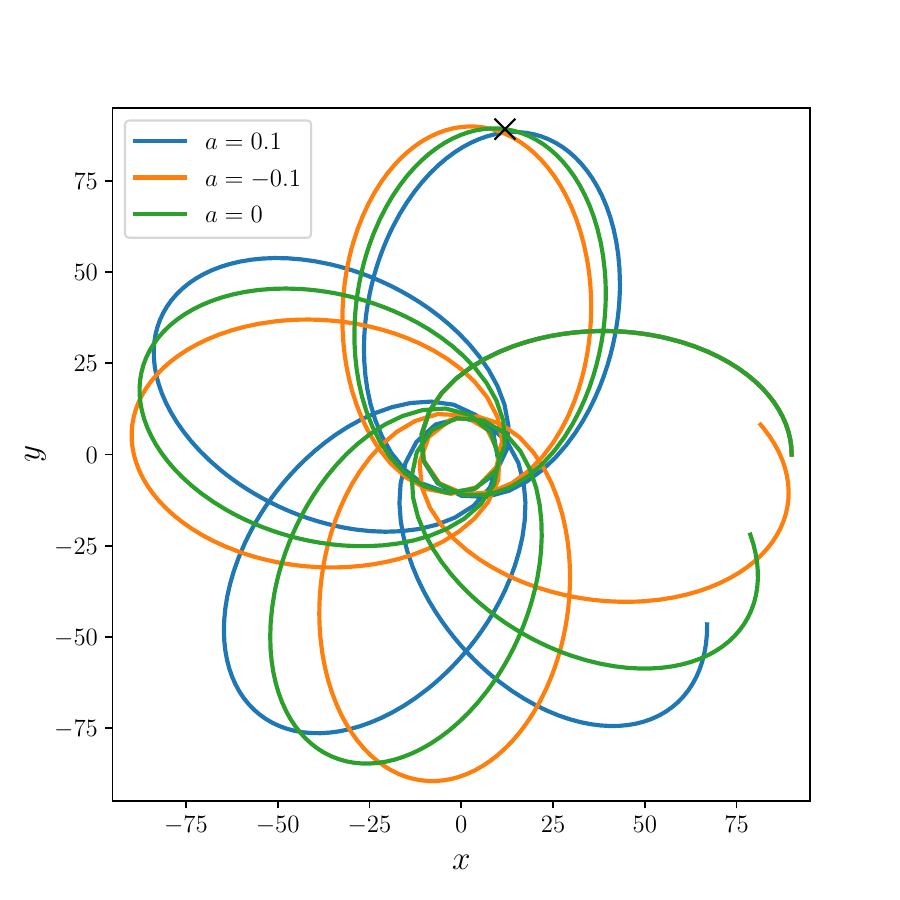}
    \caption{Free-fall dynamics in the Schwarzchild metric, starting from 12 different initial distances from the black hole in the range $[3;30]$.}
    \label{fig:precession}
\end{figure}

In order to test the ability of our code to accurately resolve the apsidal precession induced by GR onto elliptic orbits, we have evolved 32 test particles for $t=30000M$, in order to ensure the apocenter is reached at least 4 times. We  have placed our particle at the apocenter of an elliptic orbit with $r_0=90M$ and 16 homogeneously distributed initial phases $\phi$, with an initial velocity $v_\phi= 0.0521157$, and evolved it with three different metrics: $a=0,0.1$, and $a=-0.1$. In Fig.~\ref{fig:precession} we show an example of the results for particles starting on the $x-axis$ at $t=0$ as blue, green, and orange curves respectively. The black cross corresponds to the analytical prediction for the Schwarzchild case by \citet{wegg2012}, where $\Delta \phi=82.4$ after one complete orbit. Our numerical integration gives a value $\Delta \phi_{\rm num}\approx 82.39$, in perfect agreement with the expected one. Moreover, as expected, co-rotating (counter-rotating) orbits in the spinning black hole case give a smaller (larger) precession angle.
\subsubsection{Epicyclic motion}
\begin{figure}
	\includegraphics[width=\columnwidth]{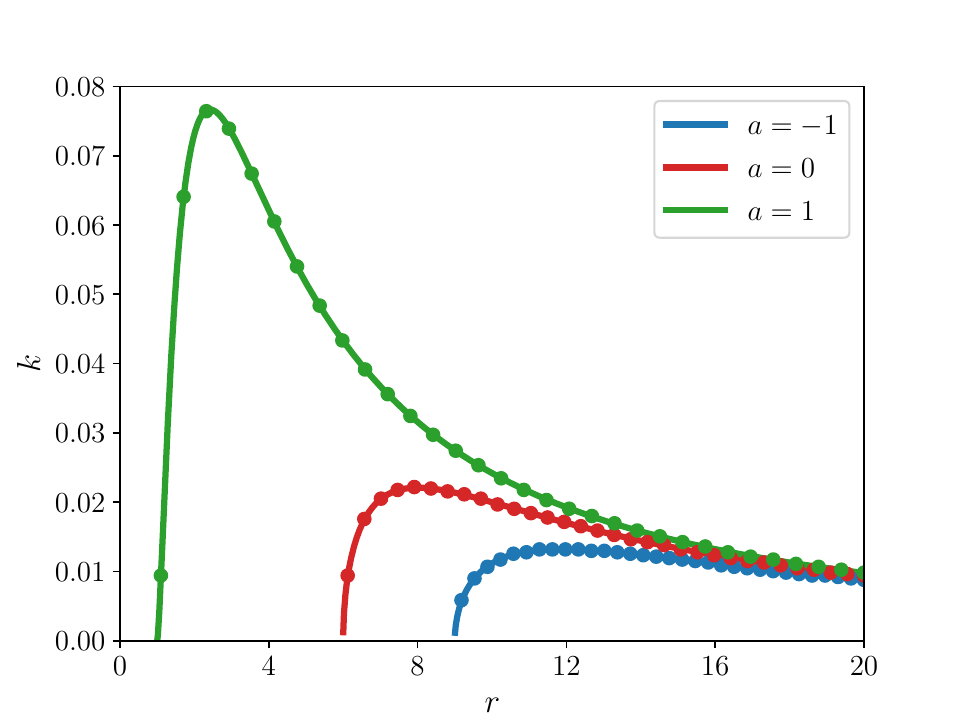}
    \caption{Epicyclic frequency $k$ for test particles in the Kerr metric as a function of radius $r$. The numerical solution for $a=-1,0,1$ is shown as coloured dots (blue, red, and green respectively), whereas the analytical result is reported as solid lines.}
    \label{fig:epicyclic}
\end{figure}

As a last test of the dynamical evolution of our scheme, we consider the epicyclic motion in the radial direction in the Kerr metric. To trigger epicyclic oscillations, we perturb the tangential velocity of an otherwise circular orbit by a factor $10^{-5}$, i.e. $v_\phi = 1.00001\Omega$. We performed three simulations, each employing 32 particles covering a radial range $r_{\rm 0} \in [1.02r_{\rm ISCO},20]$, where $r_{\rm ISCO}$ is the innermost stable circular orbit radius. As in \citet{liptai2019}, we evolve particles for $3\times 10^4$M, to guarantee that all particles have performed a few full oscillations. We then determine the epicyclic frequency $k$ using the Fast Fourier Transform implemented in the \textsc{scipy} package. The numerical results for $k$ are shown in Fig.~\ref{fig:epicyclic} for the three values of $a=-1,0,1$ as blue, red, and green dots respectively. The analytical solution is reported as solid lines using the same colour scheme, and corresponds to \citep{kato1990,lubow2002}
\begin{equation}
    k = \Omega\sqrt{1-\left(6\frac{M}{r}-8a\frac{M^{1/2}}{r^{3/2}}+3\frac{a^2}{r^2}\right)}
\end{equation}
\subsection{GR hydrodynamics: the TOV solution}
\begin{figure*}
	\includegraphics[width=0.8\textwidth,trim=2cm 1.5cm 2cm 1.5cm,clip]{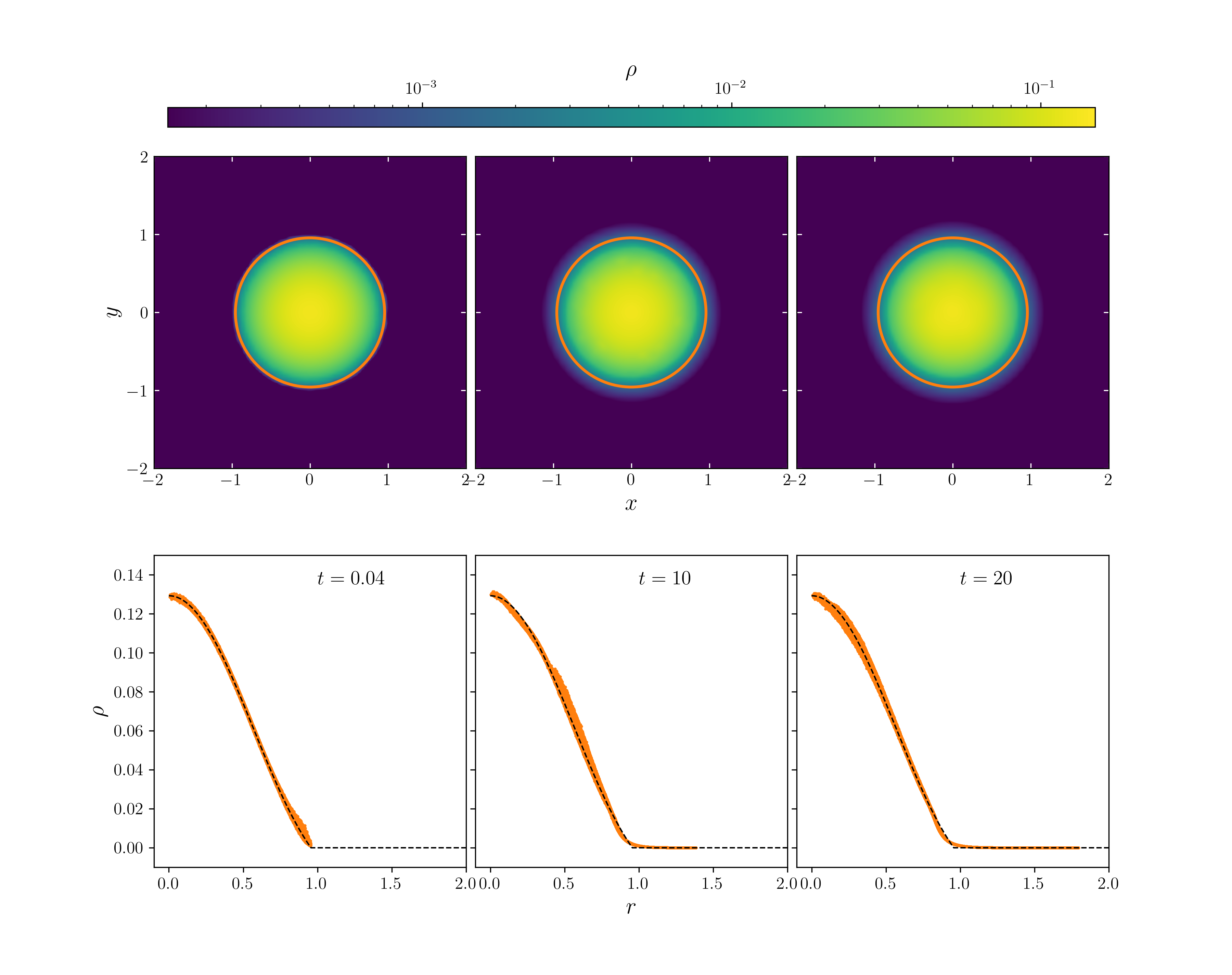}
	\includegraphics[width=0.8\textwidth,trim=2cm 1.5cm 2cm 3.2cm,clip]{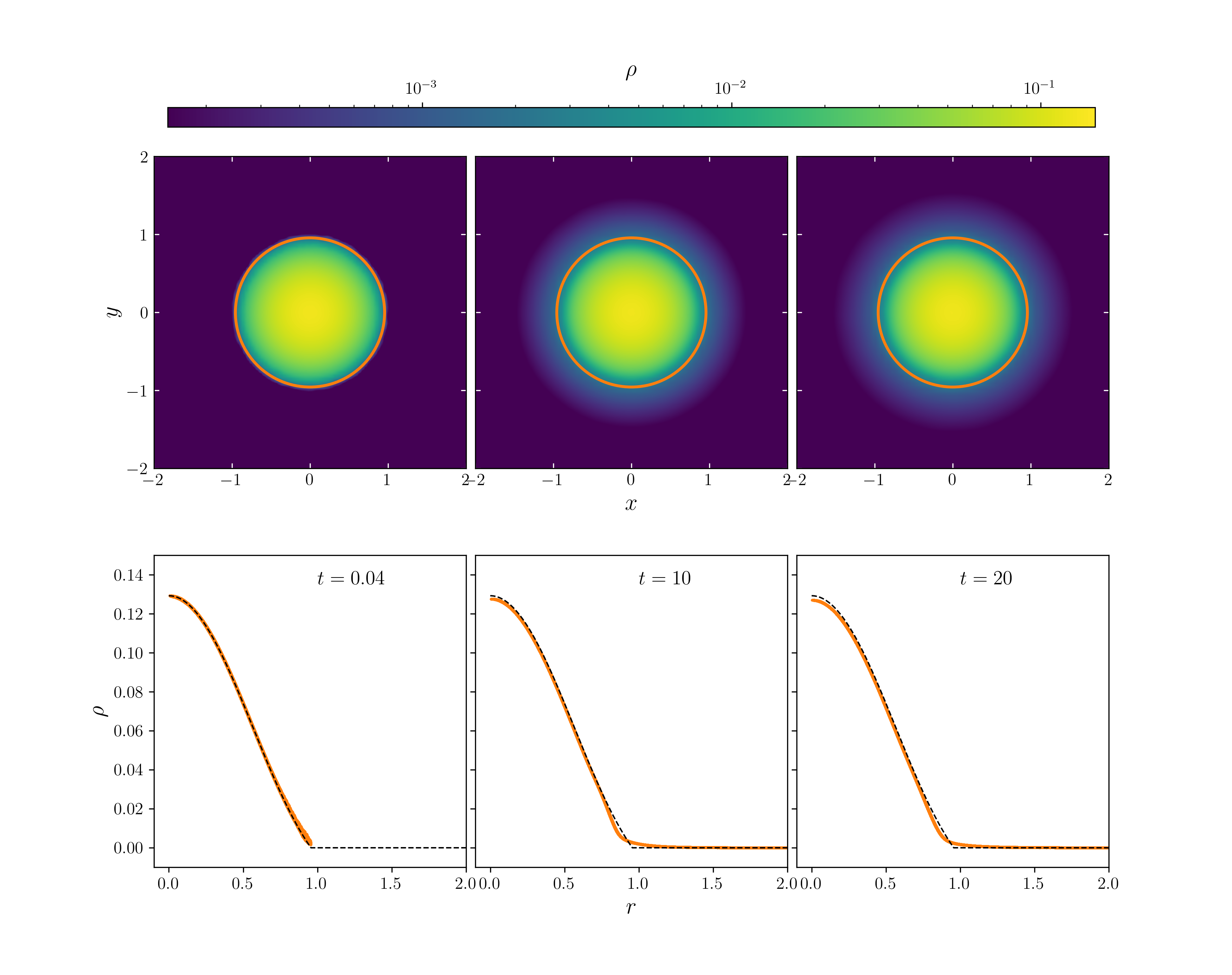}
    \caption{TOV equilibrium solution using the MFM (first two rows) and MFV (last two rows) schemes. The top panels in each pair of rows show slices of the rest mass density $\rho$ at $t=0.035t_{\rm dyn}, 10t_{\rm dyn}$, and $t=20t_{\rm dyn}$ from left to right. The orange circles correspond to the outer radius of the initial TOV solution. The bottom panels show the spherical radial profile of the numerical solution at the same times (as orange dots), compared with the exact solution reported as a black dashed line.}
    \label{fig:TOV}
\end{figure*}

In order to validate the full numerical scheme, we now consider a more general and challenging test, i.e. the stability of the Tolman-Oppenheimer-Volkoff (TOV) solution. As the TOV solution is static, any evolution in the numerical result can only be attributed to numerical errors in the calculations. Unlike in previous tests, the metric in this case is not analytic, but has to be determined and tabulated from the TOV solution. We start from a generic spherical metric written as
\begin{equation}
    ds^2 = -\exp(\nu)dt^2 + \left(1-\frac{2M}{r}\right)^{-1}dr^2 + r^2d\theta^2 + r^2\sin^2(\theta)d\phi^2,
\end{equation}
which requires us to determine $\nu$ and $M$ as a function of $r$ for the compact star.
To this aim, we numerically integrate the TOV equations
\begin{equation}
    \left\{\begin{array}{l}
         \frac{dP}{dr} = -(\mu+P)\frac{M+4\upi r^3 P}{r(r-2M)} \\
         \frac{dM}{dr} = 4\upi r^2\mu\\
         \frac{d\nu}{dr} = -\frac{2}{P+\mu}\frac{dP}{dr},
    \end{array}\right.
\end{equation}
where $P=(\Gamma-1)\rho u$ is the pressure, $\mu=\rho(1+u)$ is the mass-energy density, and $M$ is the total mass measured outside the star (notice that this does not correspond to the conserved mass we will sample with particles in the initial condition). Consistently with the assumption of cold, degenerate, relativistic matter, in the initial conditions we assume a polytropic EOS $P=\rho^2$, which translates into setting $\Gamma=2$, hence $u=\rho$.\footnote{Note that, during the evolution, we will not enforce the polytropic EOS anymore, but will use the standard $\Gamma$-law EOS.} The initial conditions are obtained by integrating the system of ordinary differential equations above from $r=0$ outwards, up to the point at which the pressure drops to $10^{-8}P_c$, where $P_c$ is the central pressure. As in \citet{chang2020}, we assume the central density $\rho_c=0.129285$. The last piece of information needed to obtain a proper solution is the boundary condition at the surface of the star $R\approx 0.9557$ for $\nu$, given by the Schwarzchild metric term (seen from outside, the star is no different from a point-like object)
\begin{equation}
    \exp[\nu(R)] = \left(1-\frac{2M(R)}{R}\right).
\end{equation}
The solution obtained is then tabulated at 20000 different radii, and passed to \textsc{gizmo} to compute all metric-dependent quantities, source terms included. 
Finally, in order to initialise our particle distribution, we first integrate the conserved mass density $D(r)$ over the radial interval $r\in [0,R]$, obtaining the conserved mass profile $\tilde{M}(r)$. Then, we distribute $\sim 10^6$ particles in a uniform density glass configuration, using the publicly available tool \textsc{WVTICs} \citep{arth2019}. We then stretch the glass configuration using the $\tilde{M}(r)$ profile, obtaining the initial conditions for our test. The left-most profiles in Fig.~\ref{fig:TOV} show the profile obtained with this procedure immediately after the beginning of the simulation, which agrees extremely well with the numerical integration result. 

We evolve the TOV star using both MFV and MFM schemes for about 20 dynamical times, with the dynamical time defined as $t_{\rm dyn}=1/\sqrt{\rho_c}$. The results are reported in Fig.~\ref{fig:TOV} after $t=0.04t_{\rm dyn}$ (left), $t=10t_{\rm dyn}$ (middle), and $t=20t_{\rm dyn}$ (right) for MFM (first two rows) and MFV (last two rows). For each snapshot, we show a slice through $z=0$ of the rest-mass density $\rho$ (top) and the corresponding radial profile (bottom).  

\begin{figure}
	\includegraphics[width=\columnwidth]{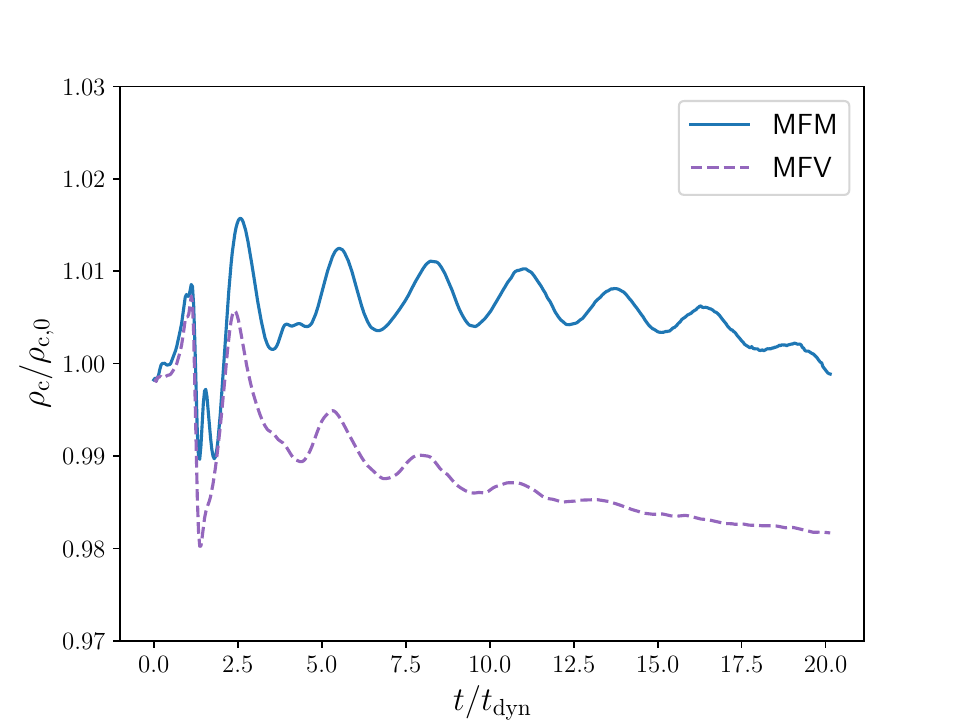}
    \caption{Evolution of the TOV central density for MFM (blue) and MFV (purple).}
    \label{fig:rho0}
\end{figure}

Both schemes are able to properly maintain the equilibrium solution over several dynamical times, despite the diffusivity of the Riemann solver and the linear reconstruction employed. In both schemes, we notice the diffusion of the star boundary outwards, enhanced by the absence of a background medium which is instead always present in moving-mesh schemes (but not strictly necessary here). This effect is slightly more severe in MFV, where the mass flux among cells tends to make gas particles more massive as they move outwards. To avoid very massive particles to form, and limit noise in the outskirts of the star, we force the code to split particles when their mass exceeded three times the initial mass. We stress, however, that this does not play any significant role in the final solution, and is not mandatory. With MFM, instead, the assumption of zero mass flux, together with the initial noise in the distribution, especially near the stellar boundary, results in a larger scatter and a moderate rearrangement of the particle distribution, as highlighted in the radial profiles, where a small overdensity is found to move inwards with time to compensate the initial diffusion outwards. Although MFV performs a bit better in this test, MFM still exhibits good convergence and stability. This is also confirmed by the evolution of the central density $\rho_{\rm c}$ over time, reported in Fig.~\ref{fig:rho0} for MFM (shown as a blue line) and MFV (shown as a purple line). Besides the initial oscillations, which occur in both schemes, MFV shows a mild decay with time, approaching 2\% at the end of the run, whereas MFM exhibits an initial enhancement by 1\%, followed by a slow decay, which approaches the initial density after 20 dynamical timescales. These small discrepancies are comparable with the results by \citet{chang2020} over similar timescales, despite the differences in the numerical schemes employed. Interestingly, however, the decay in MFV seems to flatten over time, contrary to what found by \citet{chang2020}, where the decay becomes steeper with time.

\subsection{GR hydrodynamics in the strong-field regime: the spherical accretion flow solution}
\begin{figure*}
	\includegraphics[width=\textwidth,trim=1cm 0cm 0.5cm 0.3cm,clip]{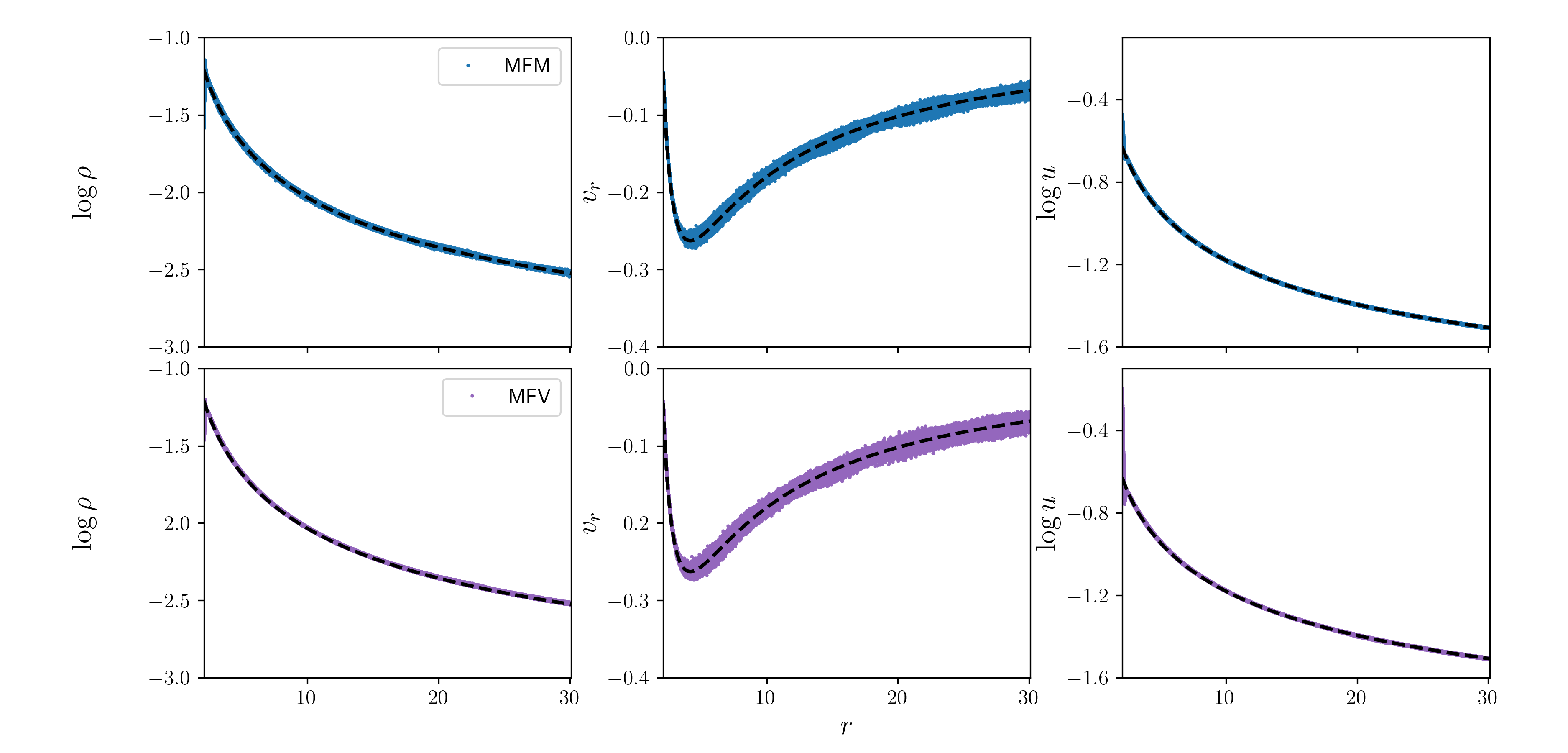}
    \caption{Spherical accretion solution in a Schwarzchild metric at $t=200$ for MFM (top panels, in blue), and MFV (bottom panels, in purple). We report the rest mass density in the left-hand panels, the radial velocity in the middle ones, and the specific internal energy in the right-hand ones.The black dashed lines correspond to the exact solution.}
    \label{fig:bondi}
\end{figure*}

Finally, we also test the ability of our code to accurately evolve hydrodynamics in the strong-field regime. In particular, we consider here a spherical accretion flow in the Schwarzchild metric of a BH with mass $M$, which is a generalisation of the Bondi solution \citep{michel1972,Hawley1984}. The solution can be written in terms of $P/\rho = (\Gamma -1)u \equiv \zeta$, as \citep[see also][]{liptai2019}
\begin{equation}
    \begin{split}
        u^r(r) &= \frac{C_1}{r^2 \zeta^n(r)} \\
        \rho(r) &= K_0\zeta^n(r)\\
        u(r) &= n\zeta(r),
    \end{split}
\end{equation}
where $n\equiv 1/(\Gamma-1)$ is the polytropic index, $K_0$ is a normalisation factor, and $C_1$ is a constant to be determined. $\zeta(r)$ can be obtained by solving the implicit equation
\begin{equation}
    C_2 = [1+(n+1)\zeta(r)]^2\left\{1-\frac{2M}{r}+[u^r(r)]^2\right\},
    \label{eq:C2}
\end{equation}
where $C_2$ is also a constant.
If we now assume a critical point $r_{\rm c}$, which gives 
\begin{equation}
    \begin{split}
        u_{\rm c} \equiv u^r(r_{\rm c}) &= \sqrt{\frac{M}{2r_{\rm c}}}\\
        v_{\rm c} \equiv v^r(r_{\rm c}) &= \sqrt{\frac{u^2_{\rm c}}{1-3u^2_{\rm c}}} \\
        \zeta_{\rm c} \equiv \zeta(r_{\rm c})& = \frac{nv_{\rm c}^2}{1+n(1-v_{\rm c}^2)-n^2v_{\rm c}^2},
    \end{split}
\end{equation}
we can determine $C_1$ and $C_2$ as
\begin{equation}
    \begin{split}
        C_1 &= u_{\rm c} r_{\rm c}^2\zeta^n_{\rm c}\\
        C_2 &= [1+(n+1)\zeta_{\rm c}]^2\left\{1-\frac{2M}{r_{\rm c}}+u_{\rm c}^2\right\}.
    \end{split}
\end{equation}
In order to define our initial conditions in terms of conserved mass $m$ and velocity $v^r(r)$, we also need 
\begin{equation}
    u^0 \equiv \frac{1}{\sqrt{-g_{\mu\nu}v^\mu v^\nu}} = \frac{\sqrt{1-2M/r+[u^r(r)]^2}}{1-2M/r},
\end{equation}
which gives
\begin{equation}
    v^r(r) = \frac{u^r(r)}{u^0}
    D(r) = \sqrt{-g}\rho u^0,
\end{equation}
where $g=-1$ for the Schwarzchild metric written in Cartesian coordinates.
In our test, we assume $\Gamma=5/3$, a critical radius $r_{\rm c}=8M$, and $K_0=1$, as done in \citet{liptai2019}, and we choose the inflowing solution in Eq.~\eqref{eq:C2}. Our initial conditions consist of an of $\sim 2\times 10^7$ particles initially placed according to a homogeneous close-packed arrangement ranging from $r=2.1M$ up to $r=100M$, which are then stretch to reflect the conserved density profile of the exact solution. During the test, particles are removed from the system as they cross the $r=2.1M$ surface. As we do not have any particle injection scheme implemented in \textsc{gizmo}, the pressure gradient at the outer edge of the sampled domain will make the gas expand at large radii, deviating from the expected solution, and this deviation will move inward with time, preventing us from reaching a proper steady-state solution. For this reason, we let the system evolve only up to $t=200$, which corresponds to the time at which the deviation appears at $r\sim 60M$. The results are reported in Fig.~\ref{fig:bondi} for MFM (top panels) and MFV (bottom panel), and show the rest mass density $\rho$ (left-hand panels), the radial velocity $v^r$ (middle panels), and the specific internal energy $u$ (right-hand panels) in the radial range $r\in [2.1M,30M]$, with the exact solution overlaid as black dashed lines. Within $r=30M$, where the fluid has had enough time to settle on the numerical solution, both schemes almost perfectly reproduce the exact profiles, with only a non negligible scatter (in particular in the radial velocity), and some small discrepancies near the horizon, where the removal of particles results in a lower density and a moderate overshooting of the internal energy. We also note that the scatter in the density profile for MFM is slightly larger than for MFV, whereas the opposite occurs for the radial velocity.

\section{Conclusions}
\label{sec:conclusions}
In this work, we have developed a GRHD extension to the MFM and MFV schemes implemented in the publicly available hydrodynamic code \textsc{gizmo}. Our implementation employs relativistic Riemann solvers to solve hydrodynamic equations, in particular HLL and HLLC (which in general is superior in accuracy to the HLL scheme commonly employed in many GRHD codes). Dynamics is instead implemented using a generalised leap-frog, as in \citet{liptai2019}, which allowed us to achieve almost perfect energy and angular momentum conservation, a unique feature of (quasi)-Lagrangian schemes compared to fixed grids. We have benchmarked our code against several standard tests, as those in \citet{marti2003} for 1D special relativistic hydrodynamics, the 3D spherical blastwave by \citet{zhang2006}, the test particle dynamics ones in \citet{liptai2019}, and \corr{challenging 3D GRHD tests as the TOV stability test \citep{chang2020} and the spherical accretion flow onto a Schwarzchild BH} \citep{liptai2019}. Despite small differences between the MFM and MFV scheme (already discussed also in \citealt{hopkins15}), both implementations showed very good accuracy in all the problems considered, and in some cases also a better convergence at lower resolution compared to other techniques. 
Our implementation, which will be made publicly available in due time, can work with any generic metric provided by the user, as we showed for the TOV stability test, although standard metrics as the Minkowski and Kerr ones are already implemented in the code. In a forthcoming paper, we will extend our implementation to include magnetic field effects, and dynamic metrics \citep[see also][]{lioutas2022}.
\section*{Acknowledgements}
AL thanks the reviewer, Daniel J. Price, for his constructive comments that improved the quality of the manuscript, and Bruno Giacomazzo and Federico Cattorini for fruitful discussions and suggestions. AL acknowledges funding from MIUR under the grant PRIN 2017-MB8AEZ.

\section*{Data Availability}
The data presented in this work will be made available upon reasonable request to the author.



\bibliographystyle{mnras}
\bibliography{biblio} 


\appendix
\section{Performance of the HLLC solver}
\label{app:HLLC}
As detailed in the main text, the HLLC solver \citep{mignone2005} is able to naturally resolve the contact discontinuity, and is only moderately more expensive than the HLL one. In order to benchmark our implementation, we report here the results of three standard tests for relativistic hydrodynamics, i.e. the 1D relativistic blast wave with and without transverse velocity, and the TOV solution. 

\subsection{Strong shocks in special relativity}
\begin{figure*}
	\includegraphics[width=\textwidth,trim=2.7cm 1.1cm 2.7cm 2cm,clip]{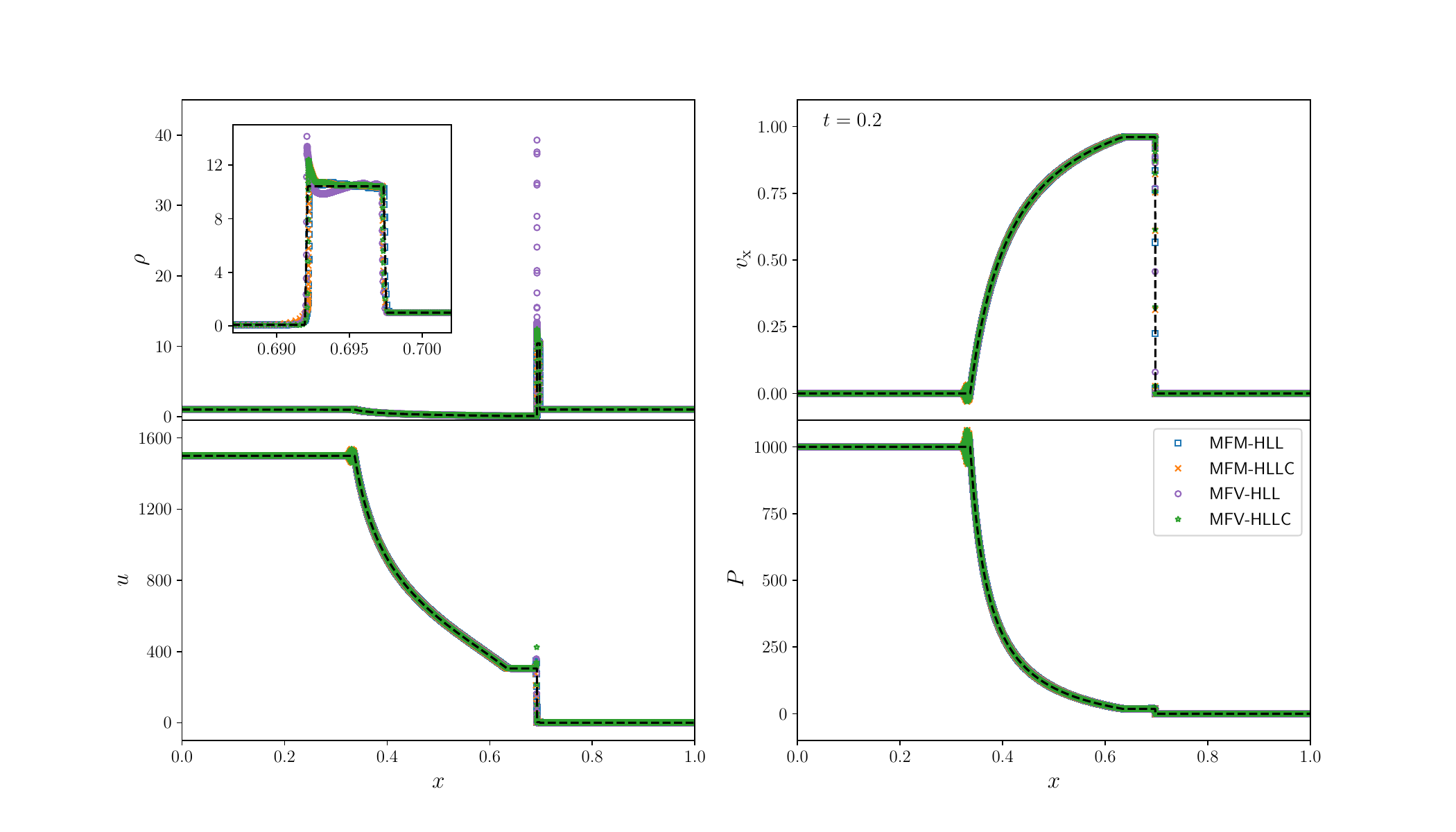}
	\caption{Same as Fig.~\ref{fig:marti2}, but comparing the high resolution runs performed with HLLC and HLL. MFM is shown as blue squares (HLL) and orange crosses (HLLC), whereas MFV is reported as purple circles (HLL) and green stars (HLLC).}
    \label{fig:marti2_HLLC}
\end{figure*}

Although not reported, we did not find any relevant difference in the use of HLL and HLLC for the mild relativistic shock reported in the main text. However, in the case of larger pressure jumps, the differences between HLL and HLLC becomes more important, as we show here. In particular, we report in Fig.~\ref{fig:marti2_HLLC} the results of the 1D relativistic blast wave test performed with HLLC. MFM (shown as orange crosses) and MFV (shown as green stars) are compared with the corresponding solution obtained using HLL (MFM shown as blue squares and MFV as purple circles). MFV is able to recover the expected solution (at the high resolution reported in the main text) similarly to HLL. Interestingly, the left side of the contact discontinuity is reproduced slightly better, whereas the right side exhibits a mild overshooting compared to the HLL solution. MFM, instead, exhibits a very large peak at the location of the density jump, exceeding the expected spike by up to a factor of four. Notice that this effect is observed only in the density, which is mostly affected by the particle arrangement, whereas the other quantities remain almost unaltered. 

\begin{figure*}
	\includegraphics[width=\textwidth,trim=2.7cm 1.1cm 2.7cm 2cm,clip]{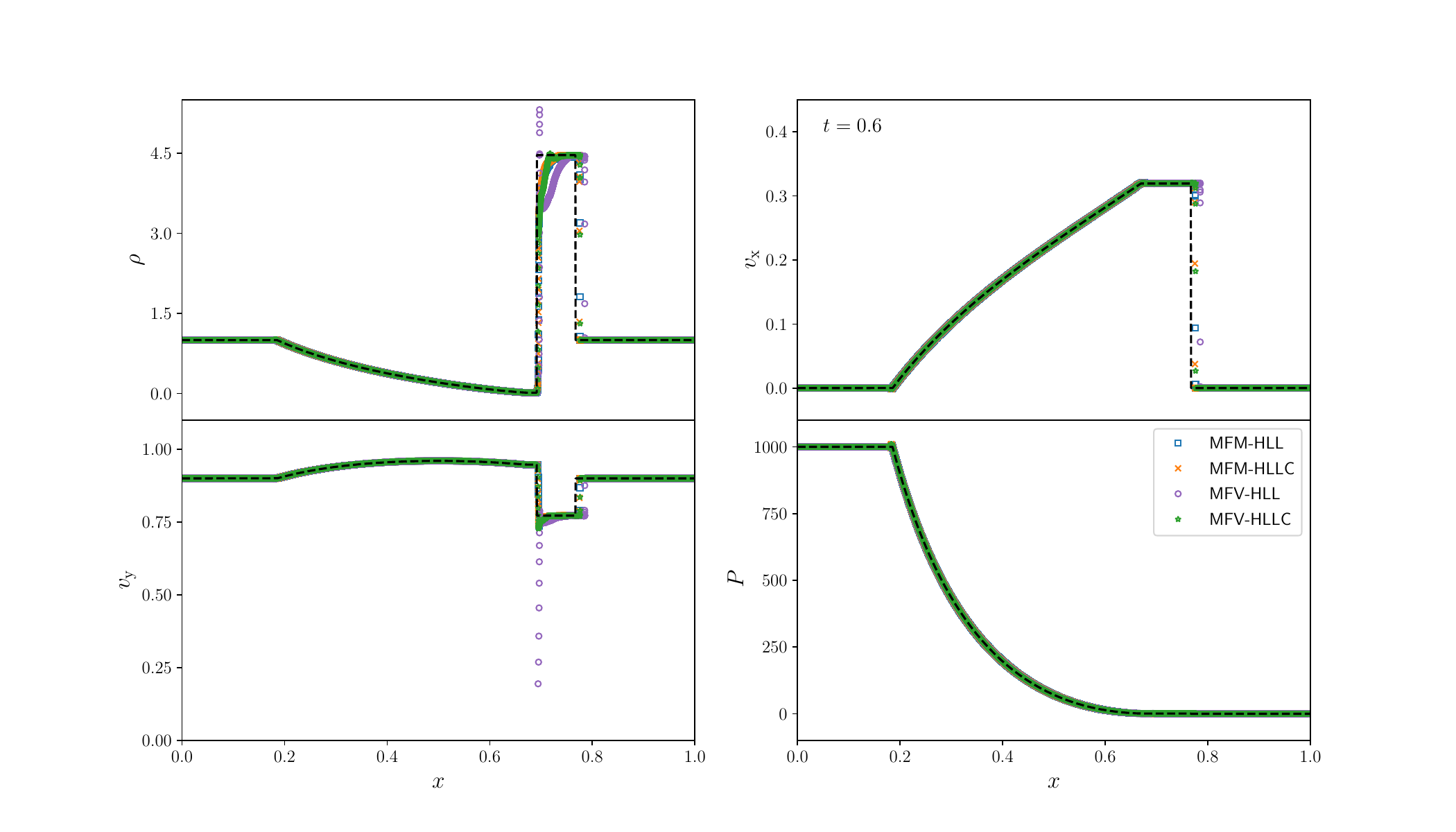}
	\caption{Same as Fig.~\ref{fig:marti2_HLLC}, for the 1D relativistic blast wave problem with transverse velocity. }
    \label{fig:marti_tr_HLLC}
\end{figure*}

We speculate this might be due to the strong coupling between gas velocities and the other thermodynamic quantities, which introduces, in the case of large discontinuities, small discrepancies between the two intermediate states, thus affecting the estimated flux. We personally verified these inconsistencies by applying the conservative to primitive solver on the intermediate states inside the Riemann solver, finding that the the recovered $v^x$ and $p$ of the two intermediate states could differ significantly from the $\lambda_*$ and $P_*$ values from \citet{mignone2005} and from each other, thus breaking the main assumptions made in the derivation. While this mild inconsistency does not significantly alter the evolution when MFV or a fixed grid are employed,\footnote{We verified this case by forcing our particles to stay fixed in space and time.} it seems to introduce numerical artefacts when MFM is used, likely because of the assumed face velocity equal to $\lambda^*$, which translates in $\bar{\mathbf{G}} = \{0,P_* \hat{n},P^*\lambda_*\}$. 

In order to test whether the presence of a transverse velocity might alleviate the inconsistencies discussed above, we also performed an equivalent run for the relativistic blast wave with a transverse velocity, which we report in Fig.~\ref{fig:marti_tr_HLLC}. Here, the MFV scheme using HLLC is almost indistinguishable from the one employing HLL, whereas MFM still exhibits a strong density jump in correspondence of the contact discontinuity (although milder than that of the previous test), followed however by a density drop inside the spike, similar to the HLL low resolution run in Fig.~\ref{fig:marti_tr}. In addition, we also notice the presence of a strong discrepancy in the transverse velocity profile, with a spike at the contact discontinuity not seen in the other cases. We note that this spike is the same observed by \citet{liptai2019} in SPH, and is likely associated to the error in the reconstruction of the velocity profile near the density jump, which according to \citet{zhang2006} requires extremely high resolution to be properly resolved. 

In general, these results seem to indicate that in the case of MFM, the HLLC solver does not perform extremely well in the presence of strong discontinuities in the fluid properties, contrary to MFV, which seems to be only moderately affected by the solver employed.

\subsection{TOV equilibrium}
\begin{figure}
	\includegraphics[width=\columnwidth]{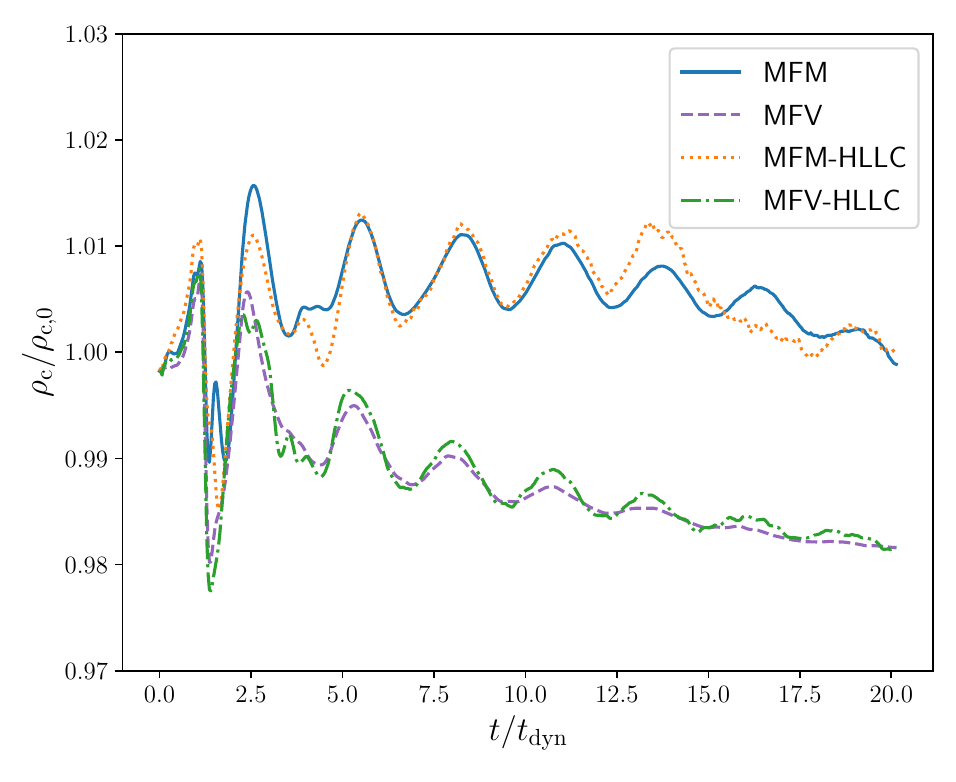}
    \caption{Evolution of the TOV central density for MFM and MFV, using the HLL solver (blue solid and purple dashed lines respectively) and the HLLC solver (orange dotted and green dot-dashed lines respectively).}
    \label{fig:rho0_HLLC}
\end{figure}

In the case of the TOV equilibrium solution, the difference between HLL and HLLC is not very important, especially in the case of MFV. This is due to the fact that, if the equilibrium is maintained with a good accuracy, the gas velocity remains very small, thus well within the regime in which the inaccuracies around the contact discontinuity have a very small effect, as can be observed in Fig.~\ref{fig:rho0_HLLC}, where we compare the central density evolution in the runs performed with HLLC with respect to those in the main text performed with HLL. MFM is shown as a blue solid line (HLL) and an orange dotted one (HLLC), whereas MFV is reported as a purple dashed one (HLL) and a green dot-dashed one (HLLC).

\bsp	
\label{lastpage}
\end{document}